# Effect of magnetic phase coexistence on spin-phonon coupling and magnetoelectric effect in polycrystalline $Sm_{0.5}Y_{0.5}Fe_{0.58}Mn_{0.42}O_3$

S. Raut[1*], S. Chakravarty[2], H.S Mohanty[3], S. Mahapatra[1], S. Bhardwaj [4], A.M. Awasthi [4], B. Kar[1], K. Singh[8], M. Chandra [4], A. Lakhani [4], V. Ganesan[4,9], M. Mishra Patidar, [4,10] R.K. Sharma[5], Velaga Srihari [6], H. K. Poswal [6], S. Mukherjee, [7] S. Giri, [7] S Panigrahi[1]

[1] Department of Physics and Astronomy, NIT Rourkela, Sundargarh-769008, Rourkela, Odisha, India

[2] *UGC-DAE Consortium for Scientific Research, Kalapakkam Node, Kokilamedu-603104, Tamilnadu, India*

[3] *Department of Basic Science and Humanities, GIET University, Gunupur, Odisha 765022, India*

[4] *UGC-DAE Consortium for Scientific Research, University Campus, Khandwa Road, Indore - 452001, Madhya Pradesh, India.*

[5] *Indus 2 Synchrotron Facility, Raja Ramanna Centre for Advanced Technology (RRCAT), Indore – 452013, Madhya Pradesh, India.*

[6] *High pressure and Synchrotron Radiation Physics Division, Bhabha Atomic Research Centre, 400085 Mumbai, Maharashtra, India.*

[7] *School of Physical Sciences, Indian Association for the Cultivation of Science, 2A&2B Raja SC Mullick Road, Jadavpur, Kolkata-32 (WB), India.*

[8] *National Institute of Technology, Jalandhar, G.T. Road, Amritsar Bypass, Jalandhar (Punjab), India – 144011.*

[9] *Medi-Caps University, Indore, AB Rd, Pigdamber, Rau, Indore, Madhya Pradesh 453331, India*

[10] *Emerald Heights International School, A.B. Road, Indore (MP) 453331, India.*

**Abstract**

The polycrystalline co-doped samples of $Sm_{0.5}Y_{0.5}Fe_{0.58}Mn_{0.42}O_3$ were prepared by solid-state reaction route and its various physical properties with their correlations have been investigated. The dc magnetization measurements on the sample revealed a weak ferromagnetic (WFM) transition at $T_N \sim 361$ K that is followed by an incomplete spin reorientation (SR) transition at $T_{SR1} \sim 348$ K. A first order magnetic transition (FOMT) around 292 K completes the spin reorientation transition and the material enters into a nearly collinear antiferromagnetic (AFM) state for T < 260 K. The compound exhibited magnetization reversal below the compensation temperature ($T_{comp}$) ~ 92 K at low measured field of 100 Oe. At further low temperature below 71 K, the compound also exhibited Zero-field cooled memory effects confirming a reentrant spinglass state formation. Robust magnetodielectric (MD) magnetoelectric coupling has been established in the present material through field dependent dielectric and resistivity measurements. *True* ferroelectric transition with a considerable value of *saturation polarization* (~ 0.06 μC/cm² at 15 K) have been found in the specimen below $T_{FE} \sim 108$ K. We observed an intense spin-phonon coupling (SPC) across $T_{SR}$ and $T_N$ from the temperature dependent Raman spectroscopy and is responsible for the intrinsic magnetoelectric effect. This SPC also stabilizes the ferroelectric state below $T_{FE}$ in the material. The delicate interplay of the lattice (Phonons), charge and spins governs the observed features in the investigated physical properties of the material that makes the specimen a promising multifunctional material.

Corresponding author: subhajitrut@gmail.com,

## I. INTRODUCTION

Temperature induced spin reorientation transitions (SR) is a phenomenon often encountered in the Rare-earth Orthoferrite ($RFeO_3$) with magnetic $R^{3+}$ ions [1, 2]. In orthoferrites, the interaction between two metallic ions sublattices namely $Fe^{3+}$-$Fe^{3+}$, $R^{3+}$-$Fe^{3+}$ and $R^{3+}$-$R^{3+}$ gives rise to a series of magnetic transitions. The dominant among these three magnetic interactions is the antiferromagnetic $Fe^{3+}$ - $Fe^{3+}$ interaction which drives the transition metal sublattice into a G-type antiferromagnetic state below $T^{Fe}_N$ (650-700 K) However, the iron spin directions are not completely collinear but are slightly canted with respect to one another. The spin canting is of two types viz. *hidden canting* causing a **C**-type or **A**-type antiferromagnetic, and the other *overt canting* causing a net moment (**F**) [1, 2]. From the symmetry considerations and the antiferromagnetic *nature* of the coupling between the magnetic ions, three types of magnetic structures are allowed in for TM sublattice in orthoferrites, $\Gamma_4, \Gamma_2$ and $\Gamma_1$. Below $T_N$ the allowed spin structure $\Gamma_4$ has the **G**-type moment directed along c-axis, while the **F** and **A** components are directed along the b and a axis of the *Pnma* crystal, giving rise to (**$G_z$, $F_y$, $A_x$**) type structure. The $\Gamma_4$ configuration can be rotated such that the net moment aligns along the *c* axis and major G-type antiferromagnetic moments lines along the b axis, then we obtain the $\Gamma_2$ configuration ($G_y$, $C_x$, $F_z$). In the $\Gamma_1$ configuration there is no net magnetization **F** along any direction and the major **G**-type antiferromagnetic vector points along the b axis (**$A_z$, $G_x$, $C_y$**).

Apart from the SR transitions, the anisotropic symmetric and antisymmetric exchange interactions between the $R^{3+}$-$Fe^{3+}$ ions can give rise to the net magnetisation reversal phenomena known also as negative magnetization (NM) within the weak ferromagnetic (WFM) order at $T^{Fe}_N$, that is not related to usual superconducting state. It rather implies that the net moment to be opposite to the applied field [5, 6]. Several of the orthoferrites and its doped samples have shown striking properties of SR transitions, NM, magnetodielectricity (MD) and magnetoelectric coupling (MEC), negative thermal expansion (NTE) [3, 5, 6, 9, 10] which renders them in a wide category of applications including sensors, thermomagnetic switches, thermally assisted magnetic random access memories, and other multifunctional devices.

Thus, a single phase material exhibiting the abovementioned properties are very rare but can be prove to be very useful for making multifunctional devices. However, in the technological point of view, it is plausible that the material become useful if the functional properties in the material occur close to room temperature or above. Several of these properties exhibited by $RFeO_3$ and their doped systems are found to be exhibited at temperatures very less than liquid $N_2$. Such as the NTE in $RFe_{0.5}Cr_{0.5}O_3$ (R= Yb, Tm) [15] and $RFe_{0.5}Cr_{0.5}O_3$ (R=Tb, Tm) [9] have been observed below the SR transitions that are < 50K. Similarly, the ferroelectricity evolved by the application of the magnetic field in $DyFeO_3$ [11] only in the coexisting short range $Dy^{3+}$ order with the WFM ordering of the $Fe^{3+}$ ions. Very few of the orthoferrites such as $SmFeO_3$ [12] and modified orthoferrites such as $YFe_{0.6}Mn_{0.4}O_3$ [3, 22] exhibits magnetodielectricity and/ ferroelectricity at RT. Recently ferroelectricity is reported also in $HoFeO_3$ [13] below $T_{FE} \ll T_N$ that is invoked by a structural phase transition to a polar *Pna2$_1$* space group from the high temperature centrosymmetric space group *Pnma*.

However, co-doped systems of orthoferrites have not been investigated so far. In this article we report about the various physical properties of polycrystalline co-doped material $Sm_{0.5}Y_{0.5}Fe_{0.58}Mn_{0.42}O_3$, that have been investigated through magnetization, dielectric, resistivity and heat capacity measurements. The dielectric and resistivity measurements under magnetic fields have also been performed in order to elucidate the presence of magnetoelectric coupling in the present system. The dc magnetic measurements showed the Néel temperature ($T_N$) of the material to be ~361 K, that is followed by an incomplete second order type spin reorientation transition ($T_{SR1}$) ~ 348 K into a nearly collinear AFM $\Gamma_1$ spin configuration. The SR1 transition into the $\Gamma_1$ phase is, however completed by a first order phase transition occurring due to the high anisotropy character of the $Mn^{3+}$ ions incorporated in the Fe sublattice. With further decreasing temperature, compensation point ($T_{comp}$) appears at ~92 K, below which the overall magnetization becomes negative. Further the ZFC memory effect study revealed a re-entrant spinglass like states coexisting with the long range ordered phase in the material for T < 70 K. The magneto-dielectric, magnetoresisitive and magneto loss measurements showed substantial linear magnetoelectric effect in the vicinity of $T_{SR2}$ and also at the low temperatures. The pyroelectric currents show occurrence of a *true* ferroelectric ordering at $T_{FE}$ ~ 108 K, within the ordered magnetic state. The temperature dependent Raman spectroscopic study revealed immense spin–phonon coupling at the magnetic and ferroelectric ordering temperatures.

## II.     Experimental Procedure

Polycrystalline $Sm_{0.5}Y_{0.5}Fe_{0.58}Mn_{0.42}O_3$ is prepared via. solid-state reaction route using high purity oxides $R_2O_3$ (Sigma Aldrich, 99.99%), $Fe_2O_3$ (Sigma Aldrich, 99.9%) and $Mn_2O_3$ (Alfa Aesar, 99.9%). Oxides $R_2O_3$ were preheated at 950 K for 7 hrs. The stoichiometric quantities of the binary oxides are then intimately mixed and heated at $1000^{O}C$ for 24 Hrs. The final calcination and sintering were done at $1350^{o}C$ for 16 Hrs several times with intermittent grinding. The single phase chemical composition is confirmed by the x-ray diffraction studies at room temperature recorded in a Rigaku x-ray diffractometer (Model: Ultima IV) using Cu K$\alpha$ radiation. Further confirmation have been done by the synchrotron x-ray diffraction studies measured at the INDUS 2, BL-11, Raja Ramanna Centre for Advanced Technology (RRCAT), Indore, India. The synchrotron powder diffraction data was analyzed using Rietveld refinement [16] available FULLPROF software.

The magnetic measurements were performed in a SQUID-VSM magnetometer (Model: MPMS 3, Quantum Design make) installed in UGC DAE CSR, Kalapakkam Node between 2-400 K. X-ray Photoemission Spectroscopic measurements (XPS) was performed at BL-14, Indus-2 with synchrotron source in ultrahigh vacuum and at ambient temperature, details of which are given elsewhere [17]. The sintered pellets were used for dielectric and impedance spectroscopic measurements in between 90-400 K using a homemade insert coupled with a Keysight E4980A LCR-meter operating at frequency range f = 20 Hz–2 MHz. The complex dielectric measurements with Magnetic field and at variable temperatures between 5 to 300 K were performed in an Alpha-A broadband impedance analyser from Novo Control using an Oxford Nano systems Integra 9 T magnet-cryostat.

The pyroelectric current ($I_p$) was recorded at a constant temperature sweep rate (5.0 K/min) in a PPMS II system (Quantum design) using a Keithley electrometer (model 6517B) and integrated with time for obtaining electric polarization (*P*). A poling field of 5 kV / cm was applied during cooling and short-circuited before the measurement of $I_p$ in the warming mode for the polarization measurement.

## RESULTS AND DISCUSSIONS

### A. Dc Magnetization studies.

The dc magnetization measurement of SYFM (58-42), as a function of temperature and under different external fields between 2-400 K, have been performed in zero field cooled (ZFC), field-cooled-cooling (FCC) and field-cooled-warming (FCW) protocols. Fig 2(a) illustrates the M (T) curves recorded under 100 Oe applied field. The ZFC, FCC and FCW M (T) curves rises sharply with decreasing T below $T_N$ ~361 K demonstrating occurrence of weak ferromagnetic transition similar to earlier reports on $YFe_{(1-x)}Mn_xO_3$ (x=0.4, 0.45) [3, 4]. However below $T_N$ several interesting features can be observed in SYFM (58-42) that are similar to that observed in $YFe_{0.6}Mn_{0.4}O_3$ but some are also different that have not been seen before in $RFeO_3$ and $RMnO_3$ systems neither in pure and nor in doped compounds. The following features have been observed.

1) With decreasing temperature, recorded M (T) curves in all the protocols undergoes a sharp decrease around $T_{SR2}$ ~ 292 K which is a convincing signature of the spin reorientation transition. The huge irreversibility between the FCC and FCW magnetization states confirms the metastable *nature* or first order nature of the spin reorientation transition similar to that observed in $DyFeO_3$ and $YFe_{(1-x)}Mn_xO_3$ [3, 4, 11, 23] below $T_N$., The merging of the FC curves below 360 K implies the second order nature of the magnetic transition at $T_N$. Comparing with the M(T) data of the present material with earlier reports on $YFe_{0.6}Mn_{0.4}O_3$ [3, 4], it can be concluded that SYFM (58-42) undergoes a temperature induced SR transition in**to** a nearly collinear ($\Gamma_1$)AFM state below $T_{SR2}$.

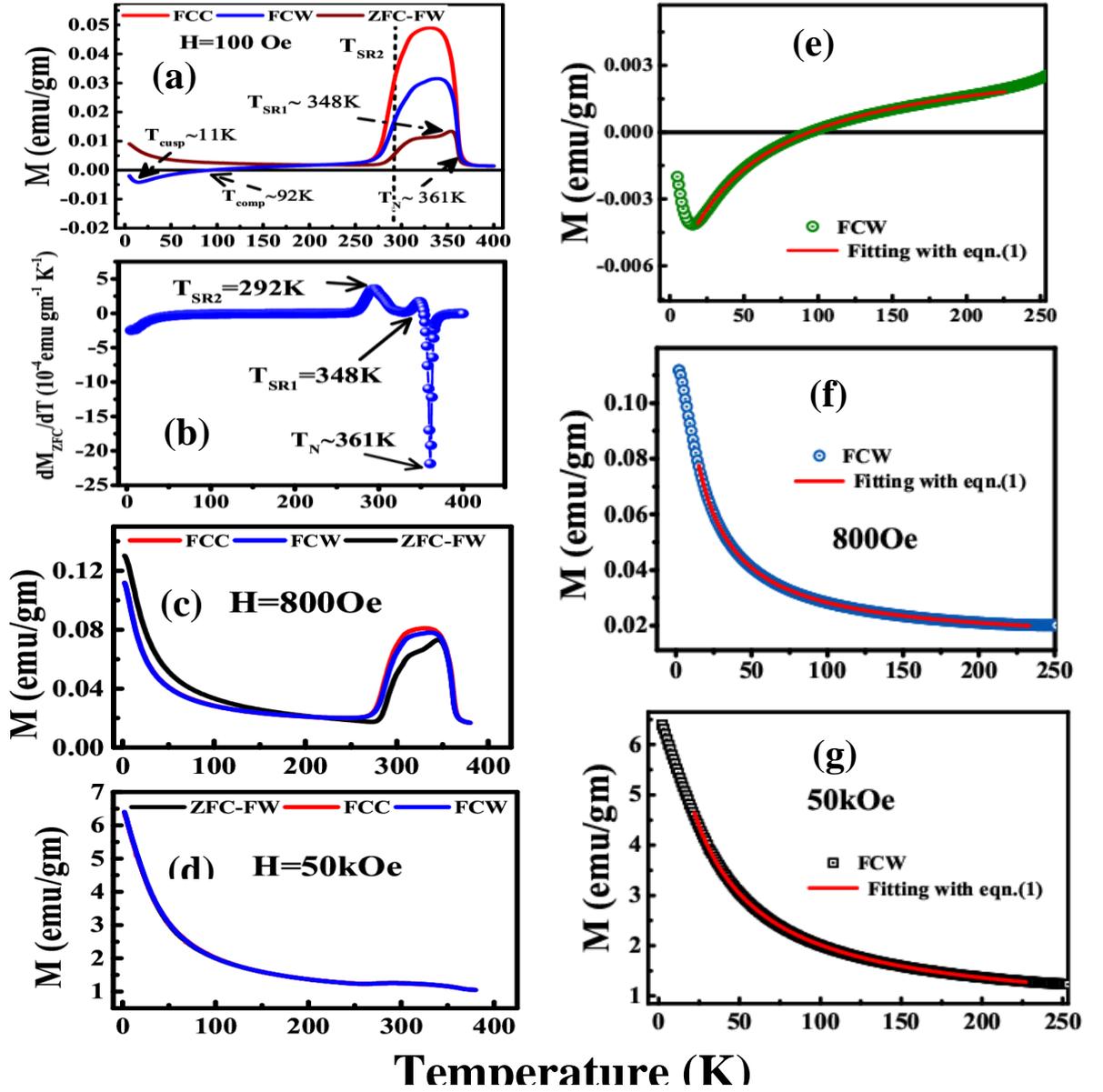

**Fig 2.** Panel (a), (c) & (d) The thermal variation of the ZFC, FCC and FCW magnetization between 2-400 K under applied field of 100 Oe, 800 Oe and 50 kOe respectively. Panel (b) T variation of the $dM_{ZFC}/dT$ showing the minimum at $T_N$ and two maxima corresponding to the inflection points at $T_{SR1}$ and $T_{SR2}$. Panel (e)-(g) shows the FCW magnetization at 100 Oe, 800 Oe and 50 kOe along with the fit with Eq (1) (solid lines).

2) It is important to note that an anomalous T variation of M(T) occurs at $T_{SR1}$ ~ 348 K as indicated in figures 2(a) and 2(b), which is more prominent in ZFC mode than in the FC modes. This anomaly can be regarded as an incomplete spin reorientation transition probably of the second order type that can be triggered by the $Sm^{3+}$-$Fe^{3+}$ anisotropic interaction below $T_N$. In YFe$_{(1-x)}$Mn$_x$O$_3$ (x< 0.45), high *anisotropic* character of the $Mn^{3+}$ ions causes the TM spin structure to abruptly change from $\Gamma_4 \to \Gamma_1$ configuration just below $T_N$ for x=0.4 [3,4]. Also the *anisotropy* interaction energy of $Sm^{3+}$-$Fe^{3+}$ interaction being comparable to $Fe^{3+}$-$Fe^{3+}$ interaction, also causes a gradual (second order) changes spin configuration $\Gamma_4 \rightarrow \Gamma_1$ in $SmFeO_3$ at elevated temperature ($T_{SR}$ ~480 K) [5]. In $Dy_{(1-x)}Sm_xFeO_3$ single crystals [23], it has been observed that the magnitude of the $Sm^{3+}$-$Fe^{3+}$ *anisotropy* interaction energy becomes dominant than the $Dy^{3+}$-$Fe^{3+}$ interaction energy for x > 0.2, because of which the exhibited second order spin reorientation transition for T > 50 K.

Hence, it is plausible to consider that strong $Sm^{3+}$-$Fe^{3+}$ anisotropy interaction energy is prevalent below $T_N$ in the present system, which induces the temperature dependent *second order* transition from $\Gamma_4 \rightarrow \Gamma_1$ phase below $T_{SR1}$ ~ 348 K. However from the view point of high anisotropy character possessed by $Mn^{3+}$ ions in the TM sublattice, continuous spin rotation process changes into an abrupt one at $T_{SR2}$, driving the system into the purely antiferromagnetic $\Gamma_1$ phase [3, 4]. Thus, the strong competition between the anisotropy energy of $R^{3+}$-$Fe^{3+}$ interaction and anisotropic nature of $Mn^{3+}$ ions governs the SR transitions in these codoped systems.

**3)** With decreasing temperature, FC curve shows negative values of magnetization below the compensation temperature $T_{comp}$ ~ 92 K. This indicates that a ferrimagnetic ground state occurs at low temperatures that can be attributed to the FC induced anisotropy of the polarised $Sm^{3+}$ ions in the AFM state as discussed latter. On contrary, the ZFC magnetization remained positive down to the low measured temperature. The specimen is in demagnetized state within the ZFC protocol. So ZFC M(T) contains nearly orthogonal AFM vectors with weak ferromagnetic components (FM) randomly distributed throughout the material. Thus, the FM components are compensated and only the AFM moments contribute the ZFC magnetization [1].

At further low temperatures FC curves exhibited an upturn towards M=0 axes indicating occurrence of another compensation point for T < 11 K. The increase in the M (T) towards positive field direction can be due to ordering of $Sm^{3+}$ sublattice that may occur at further temperatures below 2 K in SYFM (58-42) which is beyond the scope of this work.

In order to verify the contribution of polarized $Sm^{3+}$ ions by TM sublattice in inducing the observed NM below $T_{comp}$, we employed a quantitative approach using the formulation derived in Cooke et.al. [8]. The FCW magnetization within T range of 11-250 K and under different fields had been modeled separately using the following expression [6, 8]:

$$M_{Net} = M_{Fe/Mn} + \frac{C(H+H_I)}{(T-\theta)} \qquad \ldots\ldots\ldots\ldots (1)$$

Here, $M_{Net}$, $M_{Fe/Mn}$, C, H, $H_I$ and θ represents the net magnetization, magnetization due to the canted M-O-M (M=Fe/Cr), a Curie constant, an applied field, an internal field on $Sm^{3+}$ ions, and a Weiss temperature, respectively. Fig 2 (e)-(g) shows the satisfactory fitting of FCW M(T) with Eq. (1) (demonstrated as the solid line) for the applied field $H_{ext.}$ of 100 Oe, 800 Oe and 5 kOe respectively. The fitting results are listed in Table-I.

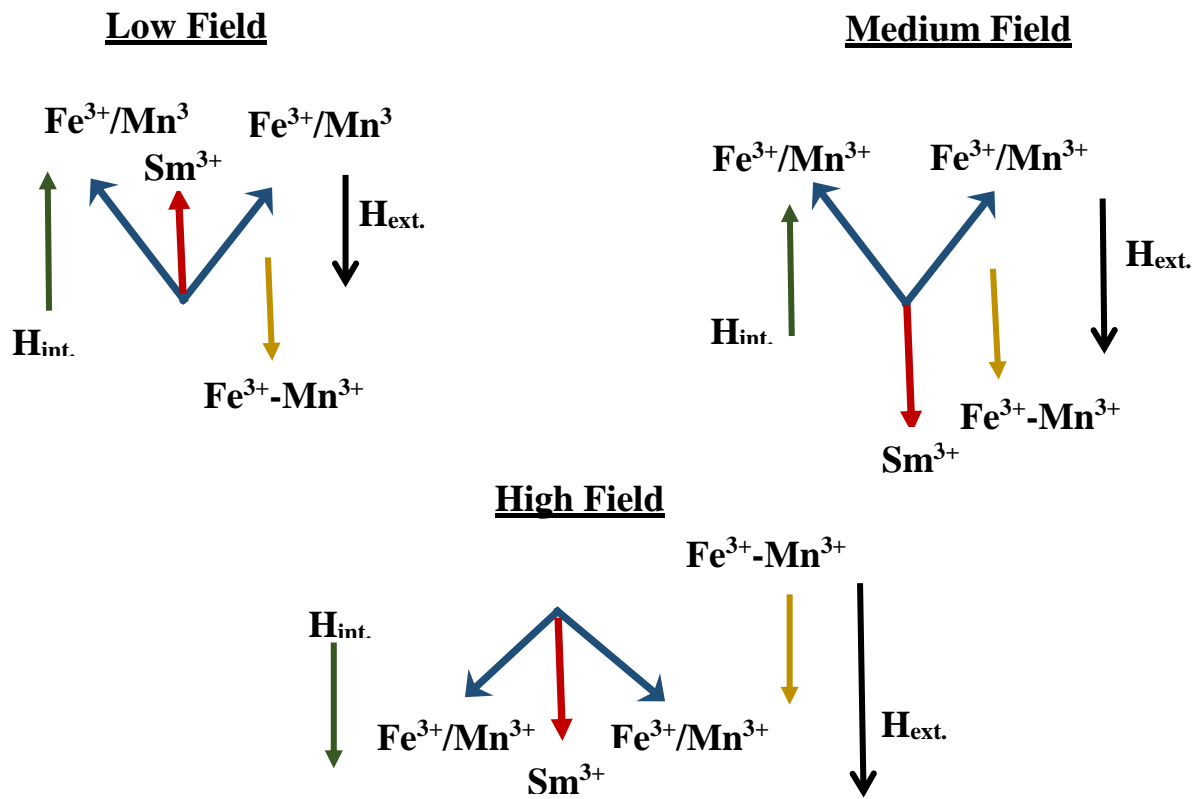

**Fig 3.** The schematic representation of the relative orientation of the Fe/Mn sublattice and the $Sm^{3+}$ ions with the external ($H_{ext.}$) and internal fields ($H_{int.}$) at different field strengths in the field cooled protocols.

As seen from Table I, the internal field $H_{int.}$ is negative. However it being slightly greater than $H_{ext.}$, the FC induced exchange anisotropy energy $Sm^{3+}$ ions from TM molecular field forces them to get aligned in the direction opposite to $H_{ext.}$ and also to the Fe-O-Mn antiferromagnetic sublattice that is oriented in the applied field direction. With decreasing T, the polarized $Sm^{3+}$ moments increase and becomes comparable with the AFM moments of the TM sublattice at $T_{comp.}$, below which the net magnetization of the sample becomes negative as seen in FC M (T) datas measured at 100 Oe (Fig 2(a)).

For $H_{ext.}$= 800 Oe, the $H_{int.}$ assumes negative values but slightly lesser than $H_{ext.}$ Thus, the $Sm^{3+}$ ions gets aligned towards $H_{ext.}$ due to enhanced Zeeman energy over the exchange anisotropy energy on the $Sm^{3+}$ ions [23]. It must be noted from Fig. (2c) that the FC-M (T) curves lie below the ZFC curve for T < 120 K. Under high magnetic fields i.e $H_{ext}$ =50 kOe, $H_{int.}$ becomes positive but less than the applied field. As a result, large moment appeared for T < 120 K as seen from figure 2(d). In addition, the first order transition is fully suppressed in higher magnetic fields (Fig 2(d)). The relative orientations of different moments relative to the internal and external fields are schematically depicted in Fig 3. This schematic representation has been represented under low, medium and high positive $H_{ext}$ respectively in Fig 3 (a)-(c). In Fig 3(a)-(c), the canted AFM moments from the Fe-Fe and Mn-Mn nearest neighbor interactions has also been shown as blue arrows that orders in the opposite field directions when cooled under low field through $T_N$.

**Table I:** The fitted parameters from Eq. (1) in M-T curves recorded in FCW modes.

| External Field (Oe) | $M_{Fe-Mn}$ (emu/gm) | $H_{int.}$ (Oe) | θ (K) | C |
|---|---|---|---|---|
| 100 | 0.0038 ± 2.139×10$^{-5}$ | -114.929 ± 0.1593 | -51.311 ± 0.6527 | 0.0373 |
| 800 | 0.0129 ± 1.753×10$^{-5}$ | -795.2576 ± 0.0078 | -11.308 ± 0.0473 | 0.3602 |
| 50000 | 0.5746 ± 0.00354 | 29964.2496 ± 333.66 | -20.7214 ± 0.1745 | 0.0022 |

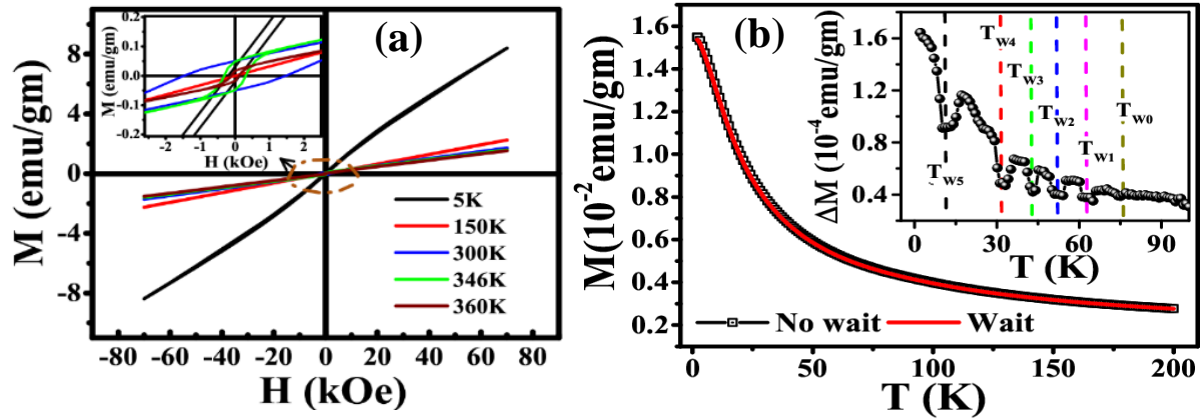

**Fig 4.** (a) Isothermal Hysteresis loop measurements at different temperatures. Inset of the figure highlights the low field region of the loops. (b) dc memory effect experiments performed without halt and with halt at three different temperatures, $T_W$ = 74, 62, 52, 43, 31 and 11 K that are indicated as $T_{W0}$, $T_{W1}$, $T_{W2}$, $T_{W3}$, $T_{W4}$ and $T_{W5}$ respectively, in the zero-field-cooled mode. The inset shows the difference in magnetization ΔM (= M without halt − M with halt) between without halt and with halt ZFC magnetization.

The zero field cooled (ZFC) isothermal hysteresis M-H loops at several temperatures, under ± 70 kOe fields are displayed in Fig 4(a). The M-H loops exhibited WFM behavior for T range 300-360 K that is expected from the M-T data [Fig 2 (a)-(d)]. The material on the other hand exhibited a linear increase in the **M** with **H** that is typical in the AFM state. At further low temperature at 5 K, a non-linear variation of **M (H)** is clearly seen, which can be understood as a complex behavior arising from the AFM (or WF) order of the Fe/Mn sublattice, superimposed to this signal, the paramagnetic response of the rare earth, which is not linear for paramagnetic $Sm^{3+}$ ions. Such nonlinearity of M-H loops below the NM has also been observed previously in the $TmFe_{0.5}Cr_{0.5}O_3$ system [15].

We have also determined the possible occurrence of a spinglass-like magnetic state (coexisting with the long range magnetically ordered state) at low temperatures which is indicated in our specific heat $C_p$ (T) data in ref (24), via. Magnetic ac susceptibility and ZFC memory effect measurements. In this protocol, the sample had been cooled from temperatures much above the assumed SG like transition $T_p = 70$ K as observed from the $C_{mag}$ (T) vs. T data [24], first in the absence of any magnetic field and without any *stop* upto 2 K. The field of 100 Oe is then applied during warming the material from 2 to 200 K, thereby constituting a *reference curve*. The sample is then cooled again form 200 K in the absence of the magnetic field but with several stops with a wait time *$t_{wait}$=5000 secs* at each stop. The temperature $T_w$ had been selected such that waiting time has been imposed on the system starting from temperatures just above $T_p$ to temperature far below it. The result of the ZFC-memory effect measurements on SYFM (58-42) is displayed in Fig 4(B). Inset of Fig 4(B) shows the difference curve $\Delta M = M_{No\ wait} - M_{Wait}$, spanning between 2-200 K. $T_{wait}$ were introduced successively at temperatures approximately 74 K, 62 K, 52 K, 43 K, 31 K and 11 K that are indicated as $T_{W0}$, $T_{W1}$, $T_{W2}$, $T_{W3}$, $T_{W4}$ and $T_{W5}$ respectively in the inset. Prominent dips can be noticed at each $T_{Wi}$ (i=1-5) i.e for $T_{wait} < T_p$. It is also observed that the magnitude of the dips increases as the temperature is decreased below $T_p$. We have also measured the *ac magnetic susceptibility* of the material ( not shown here). However, no convincing peaks around $T_p$ could be observed neither in $\chi'$(f, T) nor in $\chi''$ (f, T) vs T plots. Thus both the *dc* [(M (T) vs T] and *ac magnetic susceptibility* failed to detect the spinglass transition in the material. This can be due to strong background contribution from the long range ordered sublattice together with the increasing paramagnetic contribution of the $Sm^{3+}$ ions [25]. This observation is similar to that observed in $MnCr_2O_4$ [26] where the occurrence of the coexisting FIM and SG phases hinders the frequency dependency features in *ac susceptibility*. In that case, the memory effect measurements satisfactorily revealed the occurrence of SG transition in the material. Hence, the memory effects in magnetization exhibited below 70 K in SYFM (58-42), (which is also in conjunction with peak at $T_p$ in $C_{mag.}$ vs T data in ref. [24]), confirms the reentrant spinglass like transition occurring in the material. The occurrence of magnetic glassy phases can be attributed to the intrinsic disorderness created from the random occupancy of the magnetic ions *at same crystallographic site* and frustration of magnetic interaction namely AFM and FM between the magnetic species due to the mixed valency of Fe and Mn ions. The site and bond (FM and AFM) disorderness in the material can create a freezing in of the spins as observed earlier by us in YFe$_{0.9}$Cr$_{0.1}$O$_3$ [87]. These glassy magnetic phases are also coexisting with the long range ordered spins (AFM phase) at low temperatures in the specimen.

### c. Temperature dependent dielectric spectroscopic studies

The complex dielectric constant [$\varepsilon^*(f, T)$] of SYFM (58-42) have been measured within the T interval of 90-400 K with fixed frequencies and are displayed in Fig 5 (A)–(E). Fig 5(A) shows the plot of real part ($\varepsilon'_r (f, T)$) of complex dielectric constant vs T exhibits anomalous dielectric steps in two different thermal regimes above 150 K. The first $\varepsilon'_r (f, T)$ step [indicated by red arrow in Fig 5(A)]) manifest itself with low steepness. The second $\varepsilon'_r (f, T)$ anomaly begins from the end of the first anomaly but it is manifested as huge *dielectric* steps (indicated by black arrow in Fig 5 (A)) having a higher slope than the first one .The second anomalous $\varepsilon'_r (f, T)$ step spans between ~ 220-950 at measured frequency of 1 kHz. Since both the anomalies occurred around the vicinity of the high temperature magnetic transitions as displayed in Fig 2(a) (viz. $T_{SR1}$, $T_{SR2}$ and $T_N$,) it indicates that a substantial magnetoelectric effect [27] may be present in the material similar to that observed in YFe$_{0.6}$Mn$_{0.4}$O$_3$ [3]. On the other hand as both the dielectric anomalies are found to be frequency dependent i.e. the dielectric steps shift towards higher temperatures with increasing frequency of the measurement, it indicates that it must be associated with some low frequency relaxation process [35]. The $tan\delta (f, T)$ vs T plot as displayed in Fig 5(B) shows two set of peaks, that are coincident with the dielectric steps in $\varepsilon'_r (f, T)$ [Fig 5(A)]. The low and high T sets of peaks are designated as anomaly-I and anomaly-II respectively, and are found to be frequency dependent also. In Fig 5(B), the increased values of tan$\delta$ in the anomaly-II suggest an enhanced dc conductivity in the medium [29-31]. Inset of Fig 5(A) and 5(B) display the T variation of d$\varepsilon'_r (f, T)/dT$ and d$tan\delta(f, T)/dT$ plots respectively. Fig 5(C) and 5(D) displays the peak temperature ($T_p$) [d$\varepsilon'_r$/dT vs T plots] variation of the relaxation time ($\tau$), (calculated from the measured frequencies) for the anomalies I and II, respectively. In order to seek the dipolar cluster glass dynamics associated with these anomalies the plots of $\tau$ were fitted with Vogel-Fulcher's (VF) law $\tau = \tau^* exp\left(E_a/k_B(T_f - T_0)\right)$, where $E_a$ is the activation of energy of the relaxation process, $k_B$ is the Boltzmann's constant, $\tau^*$ is the characteristic relaxation time, and $T_0$ is the freezing temperature for dipolar dynamics. From the VF fit, thermal activation energy are $(E_a)_I \approx$ 154 meV along with the characteristic time $(\tau^*)_I \approx 3.75 \times 10^{-10}$ secs and $(T_0)_I$ = 59.78 (1.2) K for anomaly-I , while the fitting yielded the parameters as $(E_a)_{II} \approx 236$ meV, $(\tau^*)_{II} \approx 2.215 \times 10^{-10}$ secs and $(T_0) = 79.73\ (3.28)$ K for anomaly II.

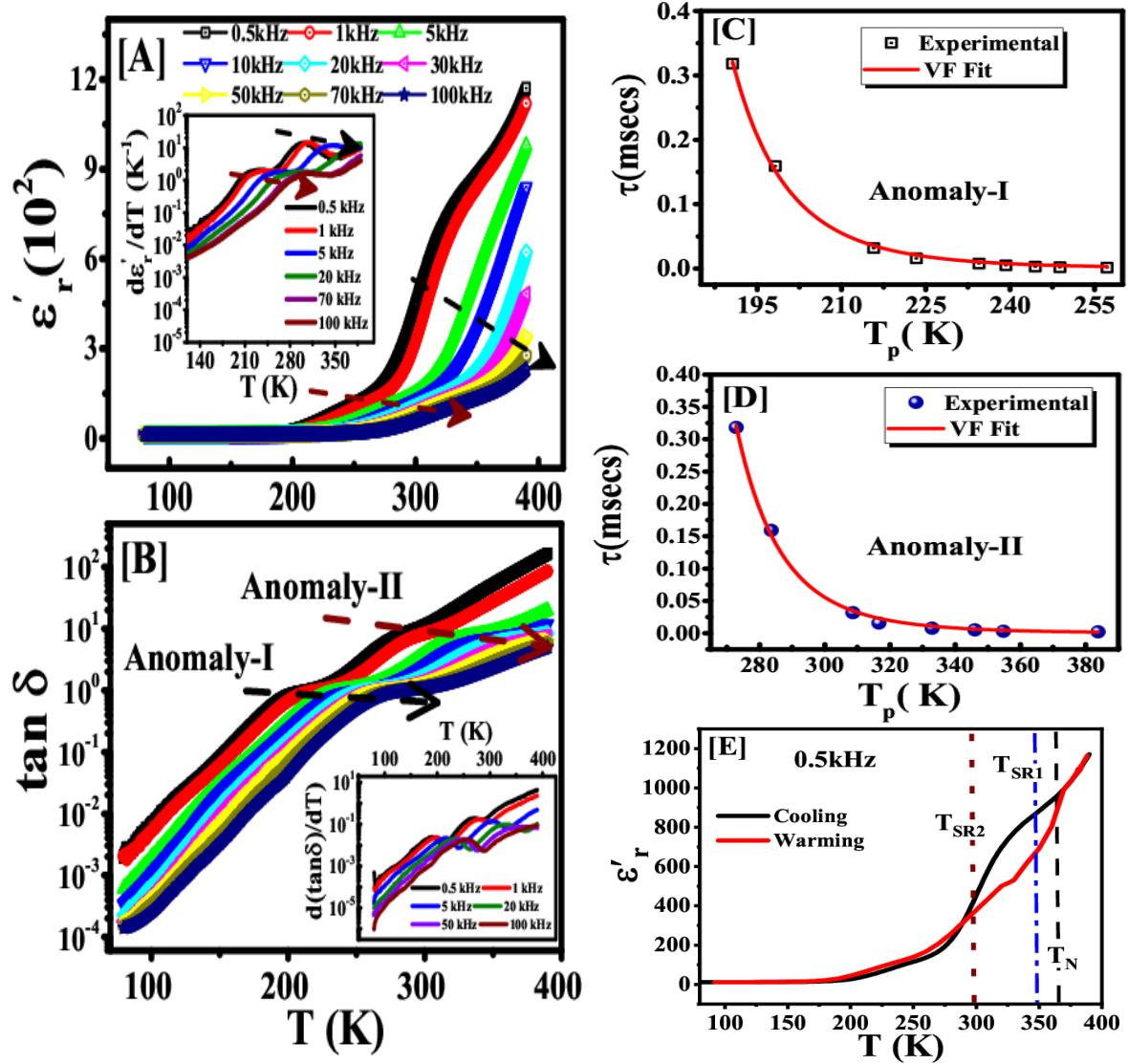

**Fig 5.** The thermal (T) variation of the real part [$\varepsilon_r'$ (f, T)] of dielectric constant and tan$\delta$ are displayed in (A) and (B) respectively at different frequencies. The insets of the figures (A) and (B) displays the respective T variation of the derivatives w.r.t temperatures. The set of anomalies I and II in the dielectric spectrum are indicated by the arrows in (A) and (B). Panels (C) and (D) displays the dependency of the relaxation times on the peak temperatures obtained from the d$\varepsilon_r'$ /d T vs. T curves (inset of (A)) for anomaly I and II respectively. The solid lines in (C) and (D) are fit with the V-F law described in the text. (E) The T variation of the $\varepsilon_r'$ (T) at 0.5 kHz in heating and cooling cycle between 90-400 K showing significant irreversibility across the FOMT envelope centered at $T_{SR2}$.

This reveals the vitreous nature of the dipolar correlations with medium-range length scale in the material. The low values of the activation energies for the relaxor dynamics of the material are of the same order as that obtained in SmFeO$_3$ nanoparticles [33], in multiglass FeTiO$_5$ [29] and also for the solid solution of the BaTi$_{1-x}$Zr$_x$O$_3$ relaxor system [34]. Moreover, the activation energy (~100 meV) is also observed in other relaxor ferroelectrics [31, 32, 34]. Fig 5(E) shows the thermal variation of the $\varepsilon_r'$ (f, T) at 0.5 kHz measured in warming and cooling cycle. A wide thermal irreversibility between the cooling and warming curves is apparent at $T_{SR2}$ also validates the first order nature of the SR2 transition. On the other hand, the reversibility between the curves at $T_N$ also establishes its second order nature.

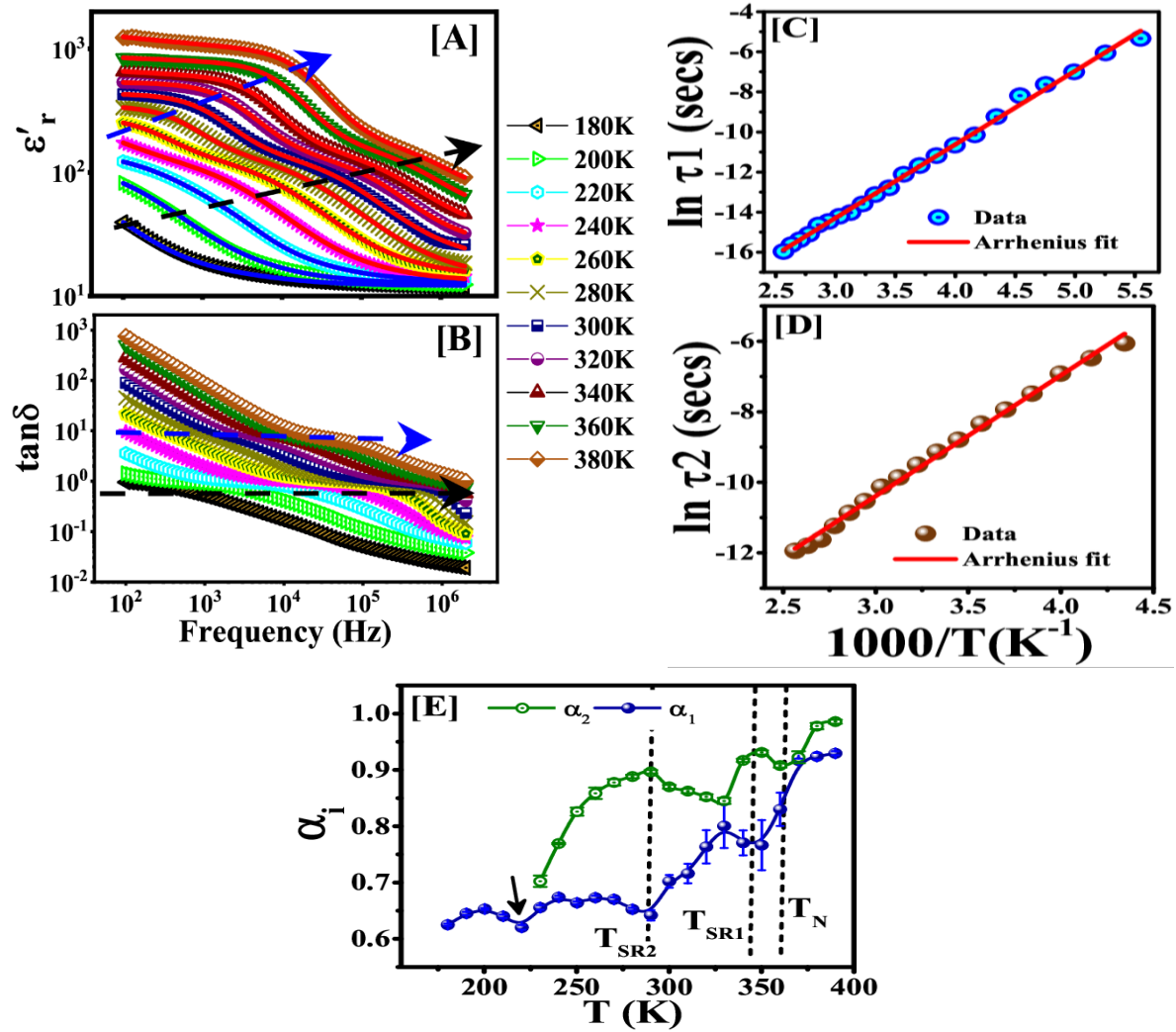

**Fig 6.** (A) and (B) shows the frequency dependence of the real part $\varepsilon'_r$ (f, T) and tanδ between 100Hz-2MHz at various selected T between 180-380 K. Joining lines are guide to the eye. Panel (C) and (D) shows the logarithm of the relaxation time (ln τ1, ln τ2) vs. 1000/T. solid lines are Arrhenius fit described in the text. (E) Shows the T variation of the parameters $\alpha_i$ (i=1, 2).

In order to seek the contribution from the extrinsic and intrinsic microstructural contribution to the observed dielectric anomalies, the dielectric and impedance spectroscopy was conducted in the temperature range of 90-400 K, within the measured frequency interval of 100 Hz-2MHz. Fig 6(A) and 6(B) displays the $\varepsilon'_r$ *(f, T)* and *tanδ (f, T)* as a function of the log *f* at different temperatures. Fig 6(A), shows that $\varepsilon'_r$ *(f, T)* vs log *f* decreases non-monotonically with a 'hump' entering from lower frequency side for T > 180 K, which typically implies the relaxation of the thermally activated defect charges (space charge/hopping polarization) originating in the grain (G) / grain-boundaries (GB) of the polycrystalline material [28]. This broad hump apparently shifted to higher frequencies with increasing T. with further increasing temperature another relaxation manifested as a nearly *f* independent '*plateau*', appeared in the spectrum from the low frequency side at T=260 K. The *f* independent '*plateau*' extends over decades of frequency with increasing T, indicating strengthening of the relaxation process [35].

The *tanδ* vs Log *f* plots also displayed humps as illustrated in Fig 6(B) that coincides with both the weak relaxation *hump* and strong relaxation *plateau* in $\varepsilon'_r$ (*f*, T) plots (Fig 6(A)).

In real dielectric materials, the dielectric relaxation cannot be modelled with Debye equation that consist single relaxation time of the polarization charges in the material and for which the $\varepsilon''_r$ - $\varepsilon'_r$ plots at particular temperature is a perfect semicircle with the origin on the $\varepsilon'_r$-axes. The polarization of the trapped mobile charges, ions as defects at grain and grain boundaries causes distribution of relaxation times that led to the formation of depressed semi circles in the argand plane of complex dielectric functions. Such non Debye relaxations can be modelled with the modified cole-cole equations [36]. According to the Cole–Cole model the dielectric functions $\varepsilon'_r$ *(f, T)* and $\varepsilon''_r$ *(f, T)* can be separately modelled as:

$$\frac{\varepsilon'_r - \varepsilon_{r\infty}}{\varepsilon_{rs} - \varepsilon_{r\infty}} = \frac{1+(\omega\tau)^{(1-\alpha)}\sin\alpha\pi/2}{1+2(\omega\tau)^{(1-\alpha)}\sin\alpha\pi/2+(\omega\tau)^{2(1-\alpha)}} \quad \ldots\ldots\ldots (2)$$

$$\frac{\varepsilon''_r}{\varepsilon_{rs} - \varepsilon_{r\infty}} = \frac{(\omega\tau)^{(1-\alpha)}\cos\alpha\pi/2}{1+2(\omega\tau)^{(1-\alpha)}\sin\alpha\pi/2+(\omega\tau)^{2(1-\alpha)}} \quad \ldots\ldots\ldots (3)$$

Here $\varepsilon_{rs}$ and $\varepsilon_{r\infty}$ are respectively the static and high frequency limits of dielectric constant, $\tau$ is the most probable relaxation time and $\alpha$ is the broadening parameter that assumes the values $0 \leq \alpha \leq 1$. The modelling of the frequency explicit plots of $\varepsilon'_r$ *(f, T) with* Eq. (2) are shown by solid lines in the Fig 6(A). However, within the entire T range, single set of parameters in Eq. (2) cannot reproduce the experimental data for T ≥ 230 K and two similar set of right hand terms of Eq. (2) were used to fit $\varepsilon'_r$ *(f, T) vs log f* plots satisfactorily (shown as the red solid lines). Solid lines in blue in Fig 6(A) represents the fit with single set of parameters of Eq. (2) *to reproduce $\varepsilon'_r$ (f, T) vs log f* data. Fig 6(C), 6(D) illustrates the thermal variation of ln ($\tau_1$) and ln ($\tau_2$) respectively and Fig 6 (E) shows T dependency of $\alpha_1$ and $\alpha_2$. Here 1 and 2 represents the high and low frequency relaxations respectively [Fig 6(A) and 6(B)]. Fig 6(C) and 6(D), demonstrated no changes in the slope either at SR or at $T_N$ of the specimen. Each ln ($\tau$) vs 1000/T plots had been treated by Arrhenius law $\tau = \tau_0 exp\left(E_a/k_bT\right)$ in order to extract the activation energy $E_a$ for the relaxations and are shown by the solid line in the figures.

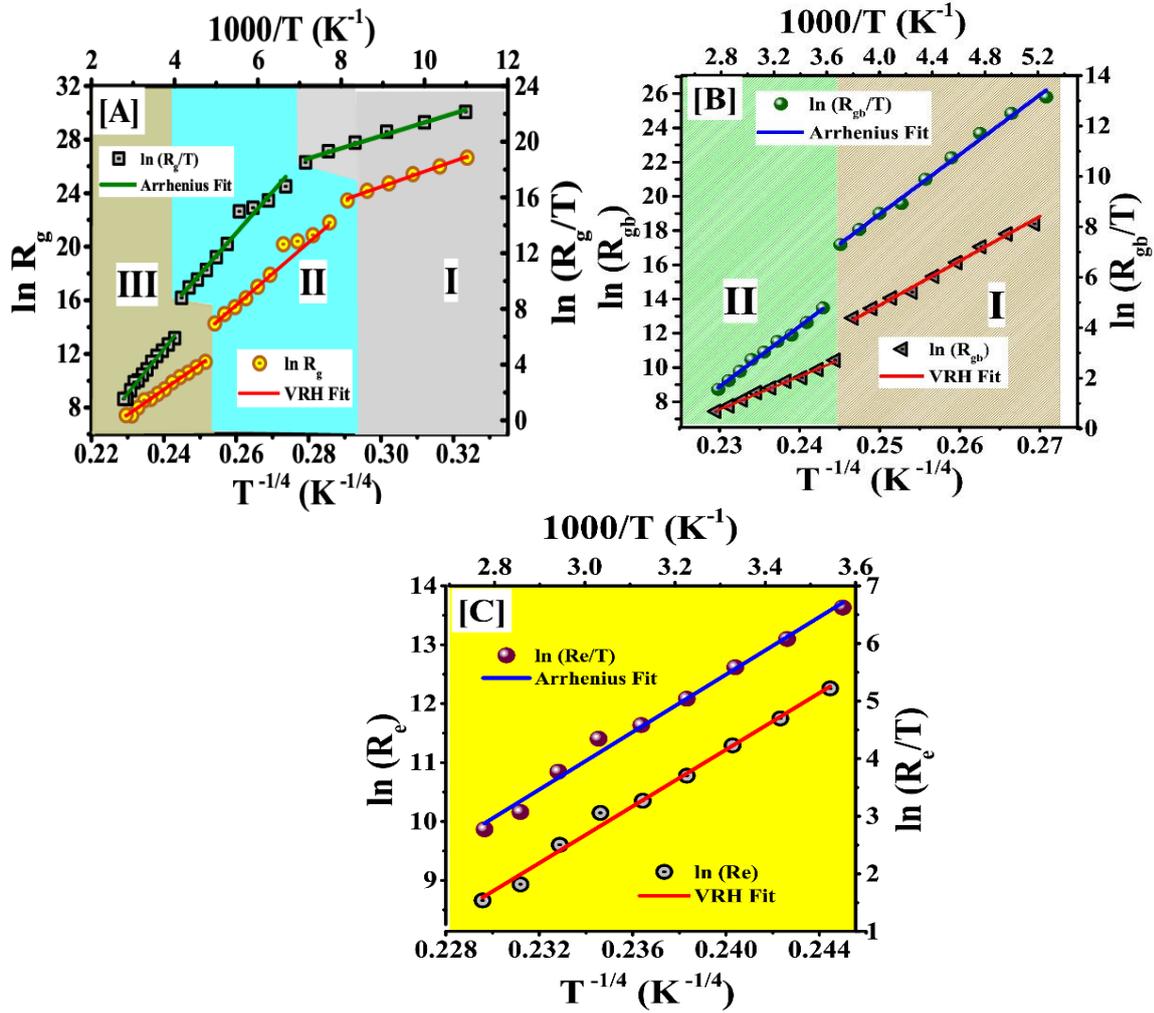

**Fig 7.** (A)-(C) shows the T variation of individual microstructural resistances $R_g$, $R_{gb}$ and $R_e$, respectively along with SPH and VRH model fits (solid lines) as described in the text.

The activation energies are calculated as $E_{a1}$=316.19 (3.8) meV and $E_{a2}$=295.31(5) meV [37]. The calculated $E_a$ values indicate the p-type small polaron hopping relaxation in the medium [38]. Fig 6(E) illustrating the T variation of $\alpha_1$ and $\alpha_2$ shows critical behaviour at $T_N$, $T_{SR1}$ and $T_{SR2}$. A dip is observed also at 220 K in the T variation of $\alpha_1$ indicated by arrow in Fig 6(E). As there are no anomalous changes in $\tau$, but there are apparent anomalies exist in the thermal variation of the width of the distribution of the relaxation times, it is suggestive that a weak coupling may exist between the magnetic and dielectric properties in the studied specimen.

Since the time constant of the relaxations are comparable it is difficult to designate them as G or GB contributions, moreover, at high temperatures electrode-material interface polarization effects (EP) may co-contribute with the other microstructural electrically inhomogeneous regions. In order to clearly distinguish the effect of the G, GB and EP effects in the *ac* electrical response of the material, the combined impedance and modulus spectroscopic analysis has been employed (See supplementary material). Quiet astonishingly, it has been seen from the combined frequency explicit plots of Z″ and M″ [Fig S3, supplementary material] that for T < 180 K only the G effect exists in the measured frequency window. For 190 K ≤ T ≤ 270 K, both G and GB effects contributes.

For 270 K < T ≤ 360 K, all the microstructural regions contribute to the measured frequency response. Hence the weak relaxation "broad hump" observed in $ε′_r$ (f, T) plot for T > 200 K [Fig 6(A)] consists of both the G and GB relaxation, while the frequency independent "*Plateau*" for T > 260 K arises from the EP effects. These observed features of co-contribution of G and GB in a single dielectric relaxation is exactly similar to that observed in CFO-PZT thick films [35]. In order to further extract the individual G, GB and EP polarization resistances and capacitances, equivalent circuit of the Nyquist plots have been employed, the details of which are given in supplementary section. The thermal variation of individual microstructural resistances $R_g$, $R_{gb}$ and $R_e$ between 90-400 K have been plotted as ln ($R_i$ /T) vs. 1000/T (i= g, gb and e) as displayed in Fig 7(A)-(C). Fig 7(A) shows the entire plot of ln $R_g$ vs 1000/T is sub-divided into different T regions I, II and III as indicated in the figure, which are having different slopes. Apparent changes in the slope of the plot occurred for T > 140 K (T region I) and for T > 220 K (T region II). It is to be noted that the slope change in T region II is also coincident with the dip in $α_1$-T variation [Fig 6(E)]. The origin of such increase in the bulk conductivity have been explained from electron-phonon interaction probed from Raman spectroscopy dc resistivity measurements in the latter section.

Fig 7(B) and 7(C) displays the thermal variation of ln $R_{gb}$ and ln $R_e$ vs 1000/T respectively. Fig 7(G) displays the abrupt changes in $R_{gb}$ leading to two distinct thermal regimes I (190 K ≤ T ≤ 270 K) and II (270 K < T ≤ 380 K) also in the plot. Each thermal regimes in Fig 7(A), 7(B) and 7(C) have been fitted with Arrhenius law, $R_i/T = R_{i0} exp (E_{act}/k_BT)$, where subscript i stands for *g, gb and e* in the expression and $R_{i0}$ is a pre-exponential factor [39]. The excellent fitting of the logarithmic resistance plots with Arrhenius law are represented as the solid lines in Fig 7 (A)-(C) also. The activation energies $E_{act.}$ for conduction, obtained from linear fitting [Fig 7(F)] are (79.71 ± 4.65) meV, (292.52 ± 19.35) meV, and (317.35 ± 10.32) meV for regions I, II and III respectively. Similarly values of $E_{act.}$ calculated from ln ($R_{gb}$ /T) vs. 1000/T plot, are (336.23 ± 9.96) meV and (340.32 ± 8.79) meV in the regions I and II respectively, while the calculated activation energy from the EP charges are obtained as (416.32 ± 15.33) meV. The obtained values of the $E_{act.}$ for conduction are similar to that obtained from dielectric relaxations in the material [Fig 6(c) and 6(d)] that suggests the same conducting charge species to be involved in dielectric relaxation in SYFM (58-42). From the above ac electrical analysis, it is clear that bulk and grain boundary conduction are dominant at low temperatures (T<280 K) for exhibiting dielectric relaxation under the *ac* field. However at the high temperatures, the EP effects begins to dominate the *ac* response of the material for T ≥ 280 K at lower frequencies, although G and GB polarization charges continues to affect the dielectric relaxations at mid and high frequencies respectively.

The linear variation of $R_g$, $R_{gb}$ and $R_e$ plots against 1000/T [Fig 7(A)-(C)] suggests the small polaron hopping (SPH) conduction of the charge carriers based on the strong electron-phonon coupling at these temperatures. Hence the plots are fitted with the Mott's SPH model [40]:

$$\rho_i = {kT\,R}/{\nu_0 e^2 c\,(1-c)}\ \exp(2\alpha R)\exp\left(\frac{W}{kT}\right) \quad\quad\ldots\ldots\ldots (4)$$

***Table II.*** *The parameters deduced from the SPH modelling and VRH modelling of the logarithm of the ln (Ri/T) and ln (Ri) versus 1000/T and 1/T$^{-1/4}$ plots respectively*

| Micro-structures | T regions | Parameters ||||||
|---|---|---|---|---|---|---|---|
| | | SPH |||| VRH ||
| | | R(Å) | α (Å$^{-1}$) | ν$_0$ (Hz) | W (meV) | T$_0$ (K) ×10$^8$ | N(E$_F$) (eV$^{-1}$cm$^{-3}$) ×10$^{20}$ |
| G | I | 5.67 | 0.298 | 3.848×10$^{12}$ | 79.711 | 0.8137 | 0.6038 |
| | II | 3.78 | 1.376 | 8.256×10$^{13}$ | 292.522 | 29.8953 | 1.6181 |
| | III | 3.78 | 1.814 | 1.002×10$^{14}$ | 317.346 | 12.6409 | 8.7676 |
| GB | I | 3.78 | 1.646 | 9.349×10$^{13}$ | 336.227 | 45.1516 | 1.8134 |
| | II | 3.78 | 1.882 | 1.029×10$^{14}$ | 340.317 | 13.9907 | 8.8464 |
| EP | I | 5.67 | 1.399 | 1.165×10$^{14}$ | 416.322 | 33.5265 | 1.5164 |

Here $\rho_i$ stands for the resistivity of the concerned microstructure, T is the temperature, *k* is the Boltzmann constant, $\nu_0$ is the optical phonon frequency, R is the average intersite separation, *e* is the electrical charge, *c* is the fraction of the transition metal ion concentration in the lower valence state namely, in our case is [Fe$^{2+}$] / [Fe$^{2+}$ + Fe$^{3+}$] or [Mn$^{3+}$] / [Mn$^{3+}$ + Mn$^{4+}$]. The fitting proceeded with α and $\nu_0$ as parameters and there values are listed in Table II. The possible intersite separation within the unit cell can be nearest Fe/Mn-Fe/Mn distances 3.78 Å or 5.67 Å which corresponds to the ion centres along b axis and in the *ab* plane along *a* direction of the unit cell. The results given in Table II are agreeable with earlier reports on similar system [18, 41]. The nonvanishing value of α indicates a non-adiabatic small polaron hopping conduction mechanism in the microstructures. However, a hopping process is of nonadiabatic small polaronic in nature, requires several restrictions on the electron transfer integral between the neighbouring hopping sites. These restrictions then serve as the criteria of judgement whether the hopping conduction mechanism is adiabatic or nonadiabatic [41, 42].

An alternative conduction mechanism [43] as demonstrated in semiconducting manganites [41] is the variable range hopping (VRH) conduction. According to the VRH model, the resistivity can be expressed as:

$$\rho(T) = \rho_0 \exp[(T_0/T)^\kappa] \quad \quad \quad \quad \ldots\ldots\ldots\ldots..(5)$$

where, $T_0$ is a characteristic temperature, $\rho_0$ is an exponential factor and $\kappa$ assumes fixed values of 1/4 or 1/2, according to the Mott regime of uncorrelated hopping carriers or for a system of carriers with a gap due to correlations according to the Efros–Shklovskii mechanism [43, 44] respectively. The plots of $\ln R_i(T)$ vs. $1/T^{-1/4}$ for G, GB and EP as displayed in Fig 7(A) -(C) (right –bottom axis) reveals excellent agreement with Eq. (5) [red lines]. The values of the characteristic temperatures $T_0$ associated with each slopes are also listed in Table II. The tabulated values shows $T_0$ of the order $\sim 10^8$ K for each microstructural regions and is agreeable with earlier literatures [41, 43]. Density of states near the fermi level $N(E_F)$ in each T regimes of G, GB and EP effects have been calculated using the relation $T_0 = 16\alpha^3/k_B N(E_F)$ and are also listed in Table II. The values of $N(E_F)$ are also found comparable to self-doped $LaMnO_3$ and similar systems [41, 47]. Thus both the SPH and NNH models can satisfactorily describe the individual T variations of the microstructural resistances in the present polycrystalline system. However due to the co-contribution of the G, GB/ EP, the overall resistance in the sample may exhibit a complex behaviour.

### d. Resistivity and Magneto resistance Measurements.

We have studied the spin-electronic correlation through measurement of *dc* resistivity $\rho(H, T)$ and magnetoresistance in the material. Fig 8 (A)-(C) shows the results of the zero-field $\rho(T)$ measurement of the specimen. The thermal variation of the resistivity of the specimen measured at zero magnetic field in the cooling and warming cycles between 150-400 K is displayed in Fig 8 (A). The system exhibited a typical behaviour of an insulator/ semiconductor within the temperature interval of 150-400 K, with a steep rise on cooling below 170 K, similar to other manganite systems [25, 48]. Below 150 K, $\rho(T)$ values are larger and so it couldn't be detected by the instrument.

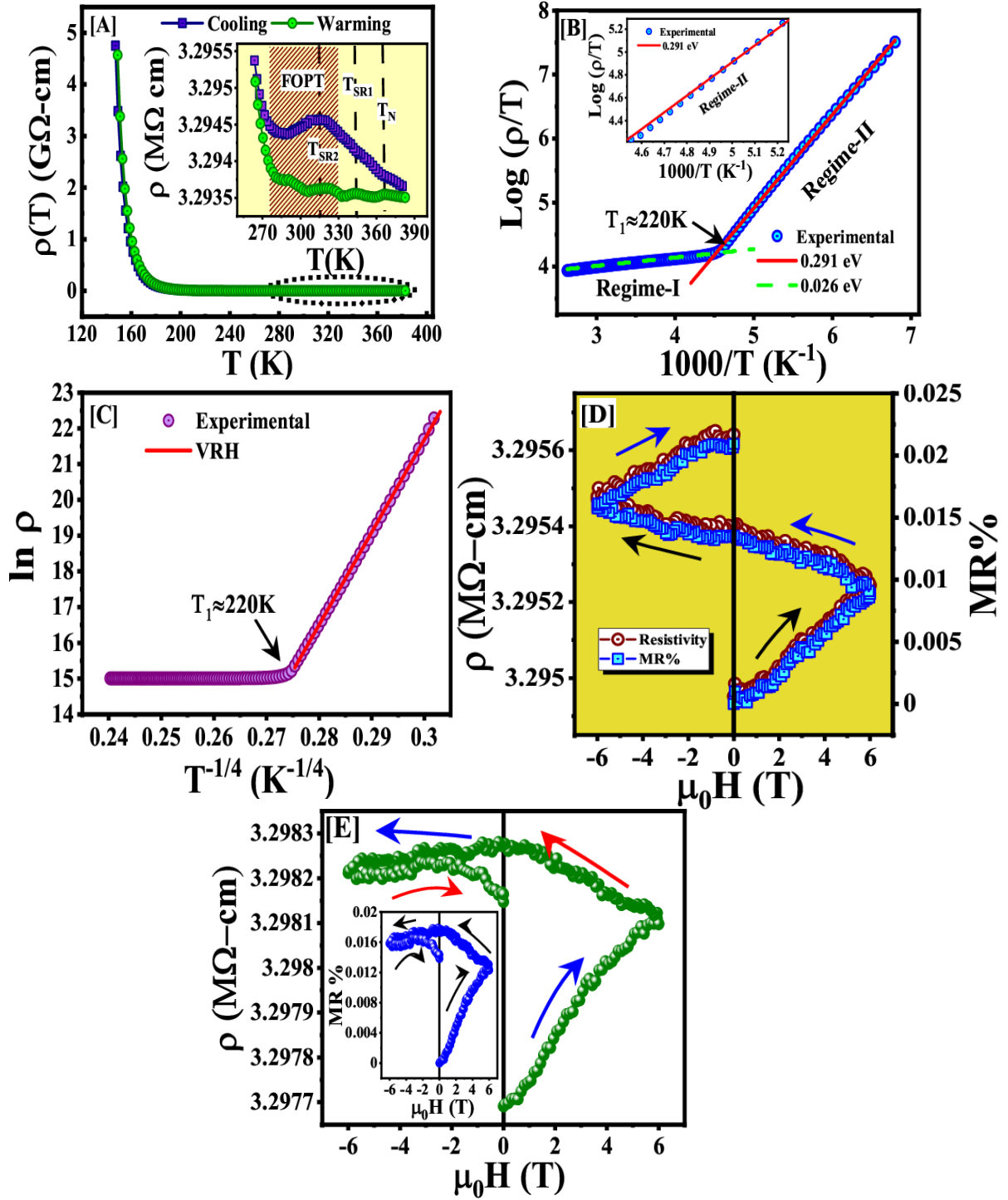

**Fig 8.** (A) The T variation of resistivity ρ(T) of $Sm_{0.5}Y_{0.5}Fe_{0.58}Mn_{0.42}O_3$ between 150-390 K in the cooling and warming cycles. Inset of the figure shows the magnified region around the magnetic transitions at high T. (B) Log (ρ(T) /T) vs. 1000/T plot and the SPH modelling with change in the activation energy at T~220 K. Inset shows the deviation of the SPH model near 220K in Regime II. (C) The ln ρ vs. $T^{-1/4}$ plot showing linear portion for T< 220 K. (D) The resistivity and magnetoresistance vs. field cycle at 300 K. (E) The resistivity vs. field cycle at 260K. Inset shows the MR% vs field variation at 260 K.

The inset of the Fig 8(A) highlights the T region with the magnetic transitions viz. SR1, SR2 and WFM transition. No sharp anomaly is vividly observed across $T_N$ and $T_{SR1}$, but a clear hump can be observed in the cooling curve across $T_{SR2}$. This implies that FOPT weakly influences electronic degrees of freedom in the material.

The sharp decrease in $\rho(T)$ for T > 170 K is due to the increase in charge carrier concentration/mobility from thermally activated carriers, at G and GB as shown in section (c), that enhances the overall $\rho(T)$ of the specimen. In order to verify the conduction mechanism for thermal activation of charge carriers, the ln $\rho(T)$ of the material has been plotted against 1000/T as displayed in Fig 8(B). The observed plot is found linear with a sharp change in the slope around T~ 220 K. Each linear T domains of Fig 8(B) viz. Regime I and Regime II, have been fitted with SPH model [Eq. (4)] using the respective slopes for calculation of the activation energies. The values extracted for $\alpha$ and $\nu_0$ are 0.1921 Å$^{-1}$ and 1.47337×10$^{13}$ Hz for Regime I with $E_{act}$ = 0.026 eV, while for Regime II with $E_{act}$ = 0.291 eV, the extracted parameter values are 1.33684 Å$^{-1}$ and 8.34461×10$^{13}$ Hz.

Comparing the results of the activation energies with that associated with the microstructural contribution to the conduction for T < 220 K, as listed in Table II, it can be concluded that dominant contribution to conduction for T < 220 K is mainly due to both intergranular (GB) and intragranular conduction (G). The change in the activation energy at 220 K with a much lower one above this temperature point indicates the change in conductivity mechanism from *p-type* to *n-type* hopping conduction [37]. We have also modelled the $\rho(T)$ data with VRH model [Eq. (5)].

As displayed in Fig 8(C), ln $\rho(T)$ vs $1/T^{-1/4}$ plot showed linearity (shown by the solid line in the figure) only for T < 220 K while VRH mechanism is failed to get obeyed for temperatures above it. Thus the VRH of the charge carriers satisfactorily explains the $\rho(T)$ behaviour for T ≤ 220 K, while the non-adiabatic SPH mechanism explains the resistivity variation for T > 220 K. It is because since the thermal energy of the charge carriers is not enough to allow electrons to hop to their nearest neighbours, the electrons favourably hop farther to find a smaller potential difference [47]. Such changes from VRH to SPH mechanism has been found in Fe doped LaMnO$_3$ [49], heterovalent doped NdFeO$_3$ [50], in mixed valent manganites [51], that can arise from electron localization due to electron-phonon (e-p) coupling. Moreover, the grain boundaries may also act as potential barriers and contribute to the localization of carriers. The e-p coupling in such mixed manganite systems can occur via three different kinds of lattice distortions:

(a) Jahn-Teller (JT) distortion of the Mn$^{3+}$ ions octahedra which raises the energy of its outermost $e_g$ electrons.

(b) Breathing-type distortion due to the presence of formally two different valence states, $Mn^{3+}$ and $Mn^{4+}$ [52].

(c) Distortion from the A-site cation size mismatch, which is valid for the present system also.

The lattice distortion arising from factors given in (a) –(c) can lead to strong e-p coupling in the specimen. The overall dc $\rho(T)$ behaviour of the material can explained by considering the carriers to be 'small polarons' as established for mixed valence manganite systems [51]. Corresponding to electron localization due to e-p coupling, the carriers are localized as small polarons with a scale of about Fe/Mn–O bond length ~2 Å [83]. At low temperatures and within the magnetically ordered state, the electrons as small polarons are self-trapped in a deep potential well. The thermal energy of the carriers being insufficient, they cannot hop out from their site, while it is more likely to be activated into an intermediate state first, which is still a localized state but with higher energy. Thus the thermal energy becomes enough for a polaron to hop to an energetically equivalent site under the influence of the magnetic localization due to spin disorder on the interatomic scale (~1nm) in the material. However the charge carriers can either hop to a NN site or to further site. This explains the satisfactory agreement of the $\rho(T)$ behaviour from both the SPH as well as VRH models in the thermal regime II [Fig 8(B) and 8(C) respectively]. With further increasing T until 220 K, the enhancement of the charge carrier concentration occurs from EP charges and also strength of electron phonon interaction gets altered. This causes further electron localization and the carriers gain sufficient thermal energy to hop to its nearest neighbouring sites without undergoing into the intermediate state. Hence $\rho(T)$ obeys the SPH model of charge carriers in regime I [Fig 8(B)]. The extracted $T_0$ of the VRH model in T regime II [Fig 8(B)] gives the value of density of states near Fermi level $N(E_F) = (0.98 \pm 0.014) \times 10^{20}$ eV$^{-1}$cm$^{-3}$ that is agreeable with earlier literatures [51, 52]. The magnetoresistances (MR) at two different temperatures 300 K (in the metastable region) and at 260 K in the AFM state as displayed in Fig 8(D) and 8(E). Both at 300 K and also at 260 K, the sample exhibited small values of MR of nearly 0.01 % at 6 T on initial field increase from 0 T to 6 T [Fig 8(D)]. Astonishingly in Fig 8(D), in the metastable region, $\rho(T)$ /MR increases linearly for applied magnetic field variation from 0 → 6 T. Since from the $\rho(T)$ measurements it is revealed that although both the AFM and WFM phases are highly resistive, yet the resistivity of the AFM state is higher than that of the WFM state. Within the metastable region, since both the AFM and WFM states coexists with minimum free energies and both have the same free energy [53-55]. Their free energy potential wells are separated by an energy barrier whose height represents the energy required for the formation of stable nuclei of the AFM phase inside the WFM phase [55, 56]. On decreasing temperature, the free energy of the AFM state becomes lower than that of the WFM as shown in Fig 9(A) for $T^* < T < T_{SR2}$. Upon crossing $T^*$, if the kinetic arrest temperature $T_g < T^*$ of the supercooled WFM phase, WFM phases transform completely into the stable AFM phase [54]. The transformation of the supercooled state can also take place even in the metastable region, upon the application of a sufficient magnetic field called the critical field $\mathbf{H_{cri.}}$ [54-57]. For several materials including Hussler alloys [53-55], doped manganites [58, 59], the application of magnetic field in the FOPT region enhances the difference between the free energy of the low T and high T states and therefore will further reduce the free energy barrier as shown in Fig 9(B). So the material undergoes a field induced transformation from supercooled high temperature phase into the low T state above $H_c$. This field induced magnetic transformation within FOPT is widely manifested as irreversibility and reversibility between the forward and backward curves for $H < H_c$ and $H \geq H_c$ respectively.

In our specimen, application of **H** in the forward and backward direction in the positive half cycle exhibits a wide irreversibility. This implies that the initial increase of the magnetic field upto 6 T is still not sufficient to convert all of the supercooled WFM phases into stable AFM phase. Upon reducing the field, the transformed AFM phase cannot go back into the metastable phase. As such the resistivity remains higher when the field is reduced to zero. It is to be noted from Fig 8(D) that ρ(T) / MR increases also during the backward paths i.e ± 6 T→ 0 T that suggests the phase transformation of the supercooled WFM state during decreasing the field, but the irreversibility between the forward and the backward process is lesser during the negative cycle than the positive one which implies that application of H > 6T can transform the metastable phases. After a complete cycle, MR increased up to ~ 0.02 % at 300 K [Fig 8(D)].

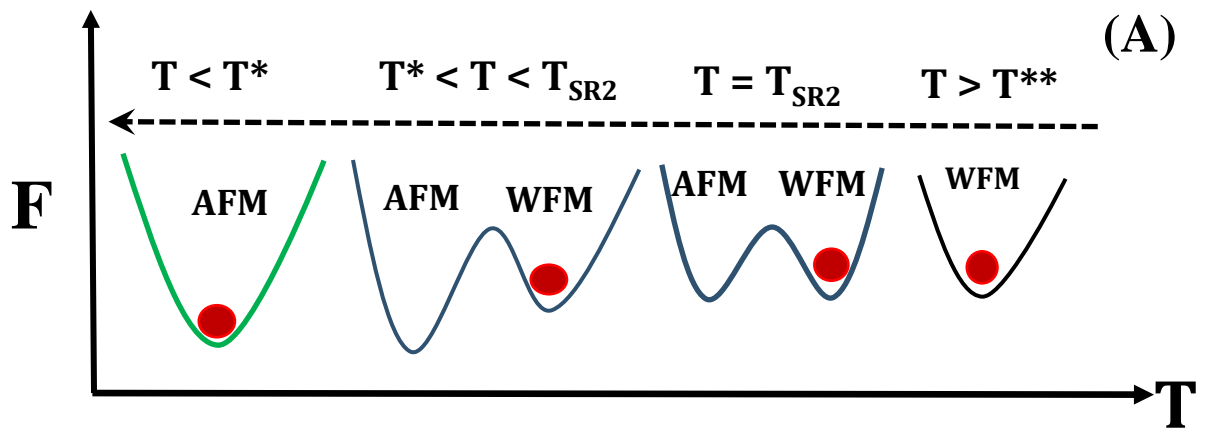

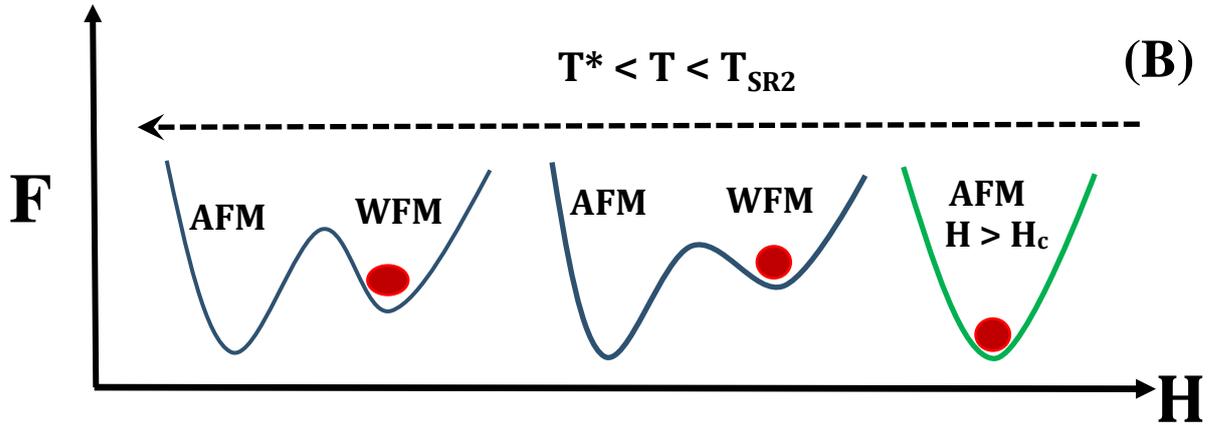

**Fig 9**. Schematic diagram of the free energies of the AFM and WFM phases across the first order phase transition. **F** denotes the free energy. The filled circles indicate the state in which the system is existing. (A) shows the variation of the magnetic phases with temperature across the first order phase transition region. Above the superheating limit T**, the WFM is the stable phase. While during cooling, the WFM phase becomes metastable in the FOPT region within the limit of supercooling T*. For T < T*, the supercooled WFM phases transform completely into the stable AFM phases if the kinetic arrest temperature $T_g$ < T*. (B) Shows the magnetic field induced transitions in the metastable region. With the enhancement of the field for H> $H_c$ the supercooled WFM phase can be transformed completely into the AFM phase.

Fig 8(E) displays ρ(T) and MR variation with **H** at 260 K. Like at 300 K both ρ(T) and MR shows huge irreversibility during the positive half cycle. This can be due to the presence of some fraction of arrested WFM phases in the stable AFM matrix. The application of the magnetic field as high as 6T is also not capable to transform the arrested WFM phases in to the low T higher resistive AFM state. While the application of the same **H** in the negative cycle seems to completely transform it. This is because due to some quenched disorderness, kinetic arrest of the FOPT lead to the some fraction of untransformed WFM phases in the AFM ground state. At T=260 K (< T*) the arrested phase being lower in concentration than the AFM phase, application of **H** transform almost entire WFM phase during the positive half cycle. Thus in negative field cycle, ρ(T) and MR remains nearly constant [Fig 8(E)]. Hence the above results of zero-field resistivity and isothermal MR measurements confirms a very weak scattering of the conduction electrons by the spins in SYFM (58-42).

E. **Magnetodielectric, pyroelectric and Raman spectroscopic Measurements**

In order to visualize directly the degree of spin-charge coupling in the specimen, the measurement of the dielectric constant and loss under variable magnetic field and at fixed temperatures had been conducted at four excitation frequencies viz. 1kHz, 10kHz, 100kHz and 1MHz. The temperature of the measurement are selected within 5-300 K. The symbols with black, red and blue colour respectively represents the datas recorded for 0→ +5T, -5T→ +5T and -5T→ 0T as shown in Fig 10(C), 10(D), 10(G) and 10(H). As shown in Fig 10(A) large changes can be observed in the $\varepsilon'_r$ (H, T) between 1 kHz to 1 MHz because of the dielectric relaxation at the concerned T. As displayed in Fig 10(C), the compound exhibited a robust magnetodielectricity at 300 K, with MD% [= {($\varepsilon'_r$ (H) - $\varepsilon'_r$ (0))/ $\varepsilon'_r$ (0)} ×100] of ~ 0.8% with initial **H** increase of 0T→+5T and for $f \geq$ 10 kHz. With further successive field branches, MD increases linearly for $f \geq$ 10 kHz in contrast to that obtained at 300 K. The Magneto loss (ML) defined as ML% [= {(tanδ (H) - tanδ (0)) / tanδ (0)} ×100] also displays a linear increases in the first increasing branch while, it decreases in the successive field branches measured at 1kHz. RT (300 K) is the temperature, where the Maxwell-Wagner (M-W) relaxation is also present along with the FOMT. Hence the observed magnetocapacitive response may comprise contributions from both MR in conjunction with the M-W effects [61, 62] (extrinsic) and also from *q* dependent spin-pair correlation function <$M_q M_{-q}$> (intrinsic) [60]. It is to be noted respectively from Fig 10(A) and 10(C) that both MD vs. H and ML vs. H for $f >$ 1 kHz mimics the MR vs H at 300K which suggest that magnetic phase coexistence and / metastabilty greatly influences the MD behaviour of the material. For MD and ML at 1kHz, the dependency on H is different from that at mid and higher frequencies, which implies that charges contributing to EP relaxation also affect the magnetodielectric behaviour of the system especially at lower frequency.

In the Maxwell–Wagner relaxation model, the real ($\varepsilon'$) and imaginary ($\varepsilon''$) parts of the dielectric permittivity are given as [62, 63]:

$$\varepsilon'(\omega) = \frac{1}{C_0(R_i+R_b)} \frac{\tau_i+\tau_b-\tau+\omega^2\tau_i\tau_b\tau}{1+\omega^2\tau^2} \quad\quad\quad\ldots\ldots\ldots\ldots (6a)$$

$$\varepsilon''(\omega) = \frac{1}{\omega C_0(R_i+R_b)} \frac{1-\omega^2\tau_i\tau_b+\omega^2(\tau_i+\tau_b)\tau}{1+\omega^2\tau^2} \quad\quad\quad\ldots\ldots\ \ldots\ldots (6b)$$

Here suffixes **i** and **b** refers to the interfacial-like (GB and EP) and bulk-like layers, respectively, *R*=resistance, *C*=capacitance, =ac frequency, $\tau_i = C_i R_i$, $\tau_b = C_b R_b$, = $(\tau_i R_b + \tau_b R_i)/(R_i + R_b)$, $C_0 = \varepsilon_0 A/t$, A=area, and t=thickness of the capacitor. Clearly, according to Eqs. (6), change in resistance of one layer invoke changes in the dielectric constant and dielectric loss in the system measured at a particular frequency also changes. Hence the combination of MR and Maxwell-Wagner (MW) effect can lead to magnetocapacitive effects. This effect of combined MR and MW is explicit to the phenomenon of true magnetoelectric effect in the material. On the other handThe intrinsic MD (that comes from the bound charges) on the other hand comes from the *q* dependency of the magnetodielectric coupling term in the free energy (F) [60]:

$$F = \frac{P^2}{2\varepsilon_0} - PE + P^2 \sum_q g(q) <M_q M_{-q}> (T). \qquad \ldots\ldots (7)$$

Here **E** is the applied electric field, ε₀ is the ''bare'' dielectric constant $g(q)$ is the wave vector dependent coupling strength in the medium, and **<M_q M_-q>** is the thermal average of the instantaneous spin-spin correlation, which obeys the sum rule:

$$\sum_q <M_q M_{-q}> = N g^2 \mu_B^2 S(S+1) \qquad \ldots\ldots (8)$$

Extremising Eq (7) w.r.t polarisation P gives $P = \frac{E}{\frac{1}{\varepsilon_0} + 2\sum_q g(q)<M_q M_{-q}>(T)} \equiv \varepsilon E$ $\qquad \ldots\ldots (9)$

Where dielectric constant $\varepsilon = \frac{\varepsilon_0}{1+2\varepsilon_0 I(T)}$, $I(T) = \sum_q g(q) <M_q M_{-q}> (T)$ $\qquad \ldots\ldots (10)$

As observed from Fig 10(C), the MD measured at 10 kHz, 100 kHz and 1 MHz frequencies increases linearly with each field branches attaining the values ~ 1.58-1.87 % respectively, at the end of the cycle. As evidenced from the Fig 10(D), the ML% vs H at 300 K exhibited following marked changes in the polarity with the field cycle as the measured *f* is varied from 1 kHz →1 MHz:

(i) The ML% at 1 kHz increases nonlinearly towards positive values with the initial **H** increase from 0→+5 T and also in the successive **H** variation from 0→+5 T. (ii) The increasing trend for 10 kHz and 100 kHz, ML during the first half cycle i.e 0→+5T→0T, is changed into a decreasing one during the second half of the field cycle i.e 0→-5T→0T.

(iii) Lastly, the ML% vs **H** plot at 1MHz becomes entirely negative for the entire **H** cycle. In order to explain the features from (i)-(iii) above, the conclusions from the dielectric and impedance spectroscopy [sec.(C)] should be recalled. The analysis of **IS** unveiled that all the three microstructural regions co-contributed to the *ac* electrical response of the sample [Fig 10 (D)] at 300K with the EP effects strictly dominating the low frequency response (*< 5 kHz*) of the material. With the enhancement of *f > 40 kHz*, the polarization from electrode-material junction as well as at the grain boundaries both relaxes while the intrinsic polarization effects from **G** remained and govern the ac response at higher frequencies (*> 40 kHz*). Indeed Catalan et. al. [61, 62] have showed that intrinsic magnetocapacitance should be measurable at frequencies higher than the conductivity cutoff (*RC* time constant). With MR% at 300 K, an order less than the MD% and ML%, it can be concluded that the observed MC and ML for *f ≥ 100 kHz* are consequences of the true ME coupling [27, 38, 60] in SYFM (58-42).

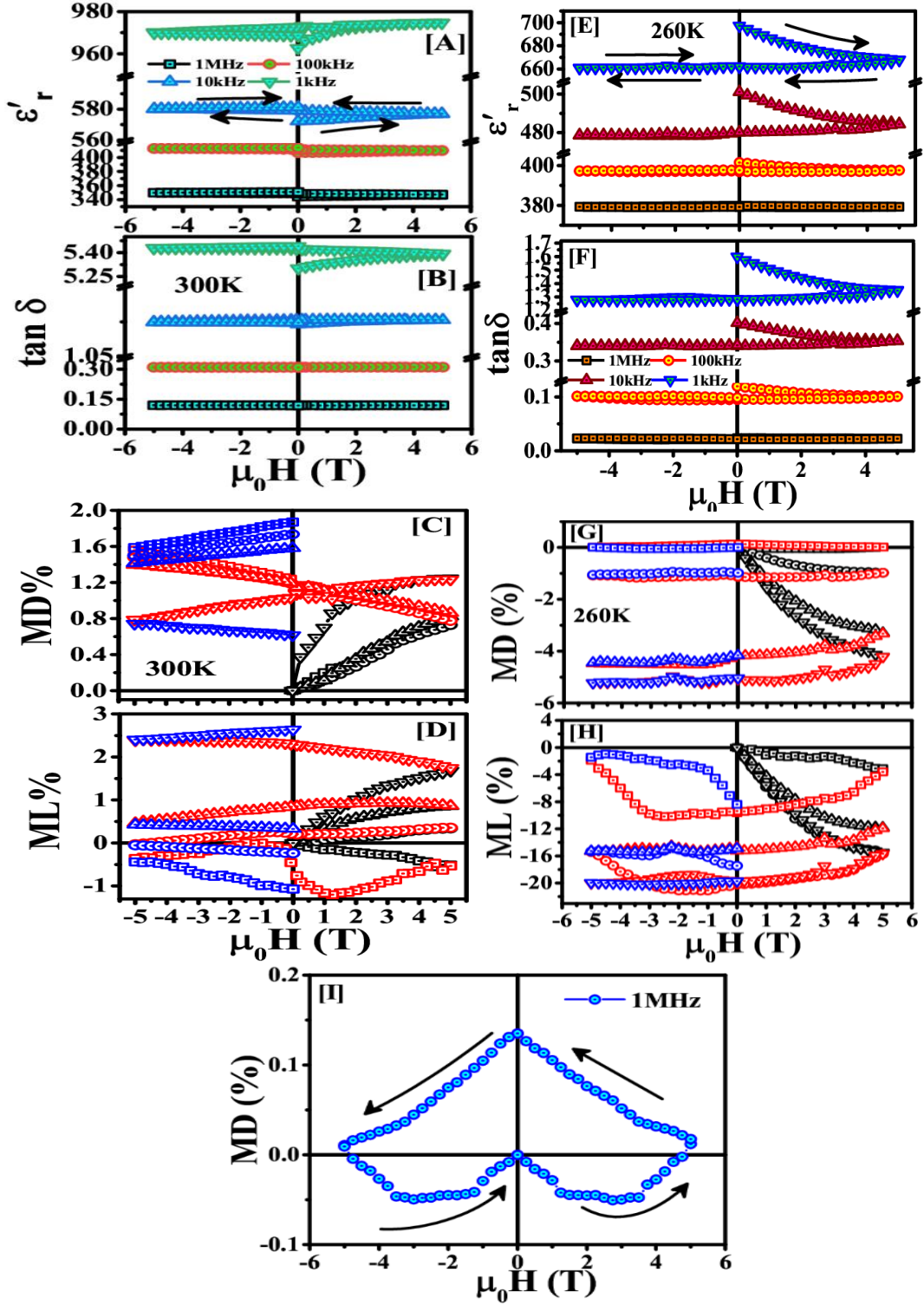

**Fig 10.** Panel (A) and (B) shows the ε′$_r$ and tanδ vs. H field variation at 300K respectively at fixed frequencies. Panel (C), (D) shows the MD% and ML% vs H for a complete field cycle at 300K. Panel (E) and (F) shows the ε′$_r$ and tanδ vs. in a field cycle at 260K at fixed frequencies. Panel (G) and (H) shows the MD % and ML % vs. in the field cycle at fixed frequencies at 260K. Panel (I) shows the MD% vs. H plot at 260K at 1MHz. For Panels (C), (D), (G) and (H) the symbols (□), (○), (△) and (▽) respectively stands for the datas measured at 1MHz, 100kHz, 10kHz and 1kHz respectively.

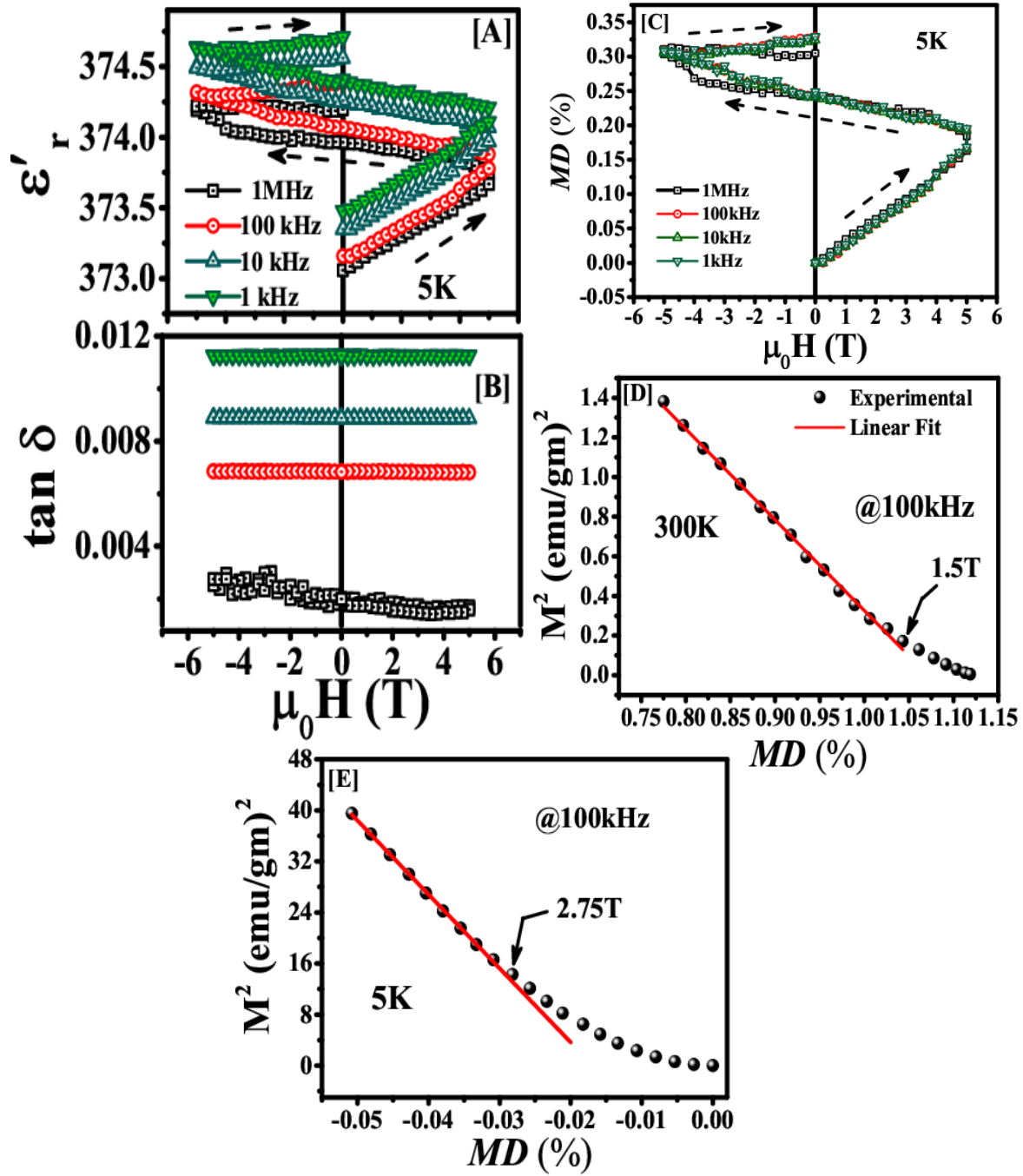

**Fig 11.** Panel (A) and (B) shows the ε′$_r$ and tanδ vs H plot respectively for the field cycle at 5K. Panel (C) shows the MD % at various frequencies for the field cycle at 5 K. Panel (D) and (E) shows the M$^2$ vs. MD % plots at 300 K and 5 K respectively at 100 kHz. The linear fit is represented as solid lines in the plots.

The intrinsic ME within FOMT should arise from the coupling of the dielectric constant and spin-pair correlation in the AFM and WFM magnetic phases. At 300 K, since the field induced transition of supercooled WFM → AFM state occurs, $<M_q M_{-q}>(T)$ also undergoes temporal and spatial variation with **H**. The resultant variation causes the MD% to increase with **H** as MR% [Fig 8(D)] in the material.

Similar to that observed at 300 K, the $\varepsilon'_r$ (T, H) changes vastly because of the presence of the relaxation as displayed Fig 10(E) and 10(F). In the entire field cycle, MD% at 260 K is found to be opposite in polarity to that at 300 K, especially for *f >1 MHz*. as displayed in Fig 10(G). Astonishingly, the MD% measured at 1, 10 and 100 kHz remained negative during the entire field cycle, while for the 1 MHz [Fig 10(I)], the MD during **H** variation from 0→5 T becomes increasingly negative attaining a saturated value of ~ -0.05% for 1.25 T < $\mu_0$H < 3.5 T. Then after, it increases with further **H** enhancement turning to positive values for $\mu_0$H > 4.75 T. With field variation +5 T→ 0 T, the MD % increases and remains positive, attaining a value of 0.135 % at the end of the positive cycle. In the second half cycle, the MD loop becomes the mirror image of the positive one resulting in a closed loop in the MD% vs. H plot. Thus in the present material both the temperature and field induced sign reversal of the magnetodielectricity occurs in the vicinity of the FOMT. The sign (negative or positive) on the magnetodielectric effect is determined by the product of spin-pair correlation of neighbouring spins and the coupling constant [38].

Since the magnetic states in the material suffer rigorous alteration when the temperature/ magnetic field is varied, both the spin-pair correlation $< M_q M_{-q} > (T)$ as well as coupling strength $g(q)$ are altered and so the magnetodielectricity changes its sign accordingly.

Fig 10(H) displays the ML of the material at 260 K in the same fixed frequencies. It can be observed that ML measured at 1 kHz, 10 kHz and 100 kHz mimics the respective MD% similar and MR% at 260 K in a field cycle. Again since the MR at 260 K is also ~ 10 times less than the MD in the field cycle, the observed MD and ML cannot be driven by the resistive components (along with MW relaxation) itself and hence must have a majority contribution from spin-pair correlation in the magnetic phases.

The results of the MD measurements at 5K are displayed in Fig 11 (A)-(C). Fig 11(A) shows a nearly frequency independent values of $\varepsilon'_r$ (T, H) due to absence of any relaxation at low temperatures. $\varepsilon'_r$ (T, H) vs H however exhibits a linear increase in the forward and backward field variation at positive and negative field cycles. The *tanδ(T, H)* assuming very small values at 5 K, exhibits field independency at all measured frequencies, as displayed in Fig 11(B).

The MD% vs **H** at 5 K is increased linearly with increasing *f* in the entire field cycle as shown in Fig 11(C) where it attains a value of ~ 0.33% at the end of the cycle. As the material is highly resistive (i.e. no relaxation effects present) for T <100 K showing negligible dielectric losses at 5 K, it can be concluded that the obtained MD% at 5 K arises entirely from the intrinsic (true) spin-charge coupling in the bulk of the material [61, 62].

An important feature to notice is the peculiar linear increase of MD% with **H** cycle at 5 K that is also observable at 260 K and 300 K. Such open loops in the MD% vs **H** plots have been previously attributed to the spurious MD signals from the charges accumulated at the grains and grain boundaries i.e in presence of M-W relaxation and MR effects [64, 65]. Although 300K is the temperature point where the MW effects are present, this increasing trend with **H** of MD% still occur at 5K where neither the MW relaxation nor the MR can confer erroneous MD signals [Fig 11(C)]. However, it is important to note that at these temperatures the appearance of the coexisting magnetic phases at 300 K (AFM/WFM) and 5 K (SG/WFIM) are common features. As stated earlier in this section, such mixed coexisting magnetic phases renders the spin- pair correlation function to vary over finite regions in *q*-space. This also results in varying coupling constant *g (q)* associated with different magnetic phases. Thus phase coexistence of WFM/ AFM at 300 K or SG/WFIM at 5 K also changes *g (q)* over finite region in *q*-space. Hence we suggest that the *q* dependent spin-spin correlation directly affects the dielectric state of the codoped system and plays a significant role in the open loop of MD% vs. **H** response against field cycling like MR% vs **H** [Fig 8(D) and 8(E)].

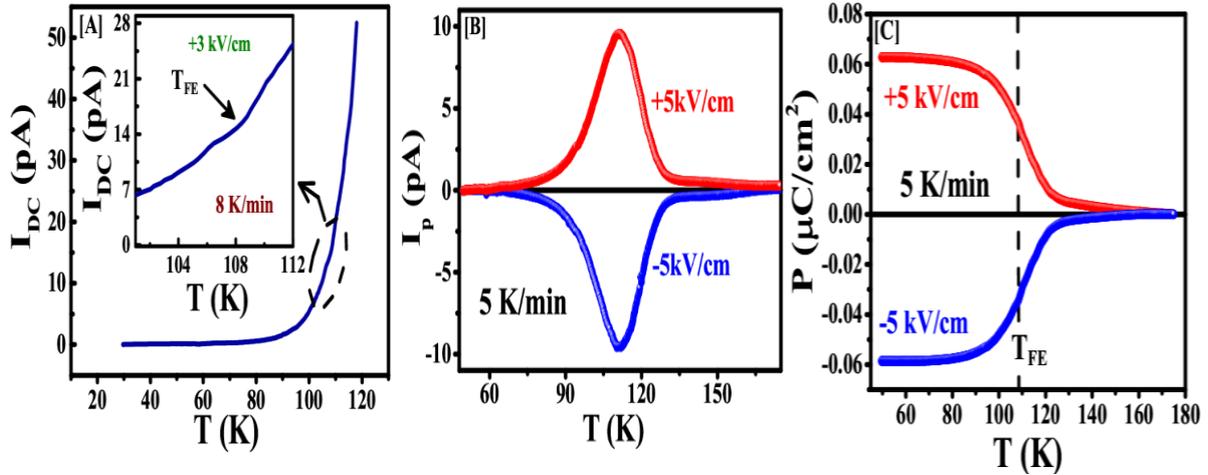

**Fig 12.** (A) Shows the bias electric field (BE) measurement showing the true ferroelectric transition at $T_{FE}$ =108 K with a temperature sweep of 8 K/min. Inset shows the zoomed portion around $T_{FE}$. (B) Shows the pyroelectric current measurements under ±5 kV/cm electric field .with T ramp of 5 K/min between 50-175 K. (C) Shows the T variation of the ferroelectric polarization (**P**) of the material between 50-175 K, under poling fields of ±5 kV / cm. with T ramp of 5 K / min.

The Intrinsic MD effect can be phenomenologically described by the simple Ginzburg-Landau theory for phase transition and is attributed to the ME coupling term $\gamma P^2 M^2$ in the thermodynamic potential ($\Phi$) given as

$$\phi = \phi_0 + \alpha P^2 + \alpha' M^2 + \frac{\beta}{2}P^4 + \frac{\beta'}{2}M^4 - PE - MH + \gamma P^2 M^2 \qquad \ldots\ldots (11)$$

Where *α, β, α′, β′,* and *γ* are the constants and functions of temperature. Thus, the influence of the magnetic order on the magnetic field driven magnetodielectric can be followed from the linear variation of the MD% vs $M^2$ curve [66-68]. The square of the magnetization $M^2$ against the MD response at different temperatures for the specimen showed linear behaviour in the high field regimes i.e for H > 1.5 T at 300 K and H > 2.75 T at 5 K as depicted in Fig 11(D) and 11(E). This proves that the linear coupling term $\gamma P^2 M^2$ term of the Ginzburg-Landau theory [Eq. (11)] is significant for SYFM similar to spinels $MCr_2O_4$ (M= Mn, Ni, Co) [67, 68]. We have also investigated the occurrence of ferroelectricity by measurement of pyroelectric current. Primarily to seek the genuine ferroelectric transition in the material, the bias electric field method (BE) had been employed as recently described by N. Terada et al. [69]. Afterwards the pyroelectric current measurement had been conducted within the T range of 50-175 K. As displayed in Fig 12(A), the current $I_{DC}$ measured in the bias electric field method under applied field of 3 kV/ cm with a T sweep rate of 8K/min reveals a slight dip around 108 K as indicated in inset of Fig 12(A). The appearance of the dip in $I_{DC}$ as a function of T [Fig 12(A)] confirms the occurrence of *true* ferroelectric state in SYFM (58/42) [96] with $T_{FE}$~ 108 K as the ferroelectric transition temperature. The pyroelectric current ($I_P$) measurements under ±5 kV/cm between 50-175 K, revealed identical ± $I_P$ peaks at T~ 108 K as displayed in Fig 12(B). Fig 12(C) displays the time integrated $I_P$ that gives intrinsic saturation electric polarization ($P_s$) values of about ± 0.06 μC/ $cm^2$ under ± 5kV/cm poling field respectively below $T_{FE}$. Reversal of *P* due to a change in sign of *E* signifies ferroelectric behaviour of SYFM (58-42). The value of the $P_s$ obtained in the SYFM (58-42) is comparable to several improper ferroelectrics [67-69] and hence confirms the involvement of long-range ordering of electric polarization.

To seek the intrinsic origin of the magnetodielectricity and ME coupling, Raman spectroscopic measurements at variable temperatures had been conducted inside the T interval of 83-503 K. The dielectric constant of a material usually depends on the long wavelength longitudinal and transverse optic phonon frequencies through the Lydian-Sachs relation. Hence, affecting the phonon mode at the magnetic transition through spin-phonon coupling, alters the dielectric constant of the material thereby giving rise to phonon mediated ME effects that is intrinsic in nature. Indeed, for isostructural $SeCuO_3$ and $TeCuO_3$ it has been grounded both theoretically and experimentally that coupling of the spin fluctuations affecting the optical phonons can give rise to magnetodielectric effects [60]. Unpolarised Raman spectra of SYFM (58-42) are displayed in Fig 13(A) at selected temperatures. The spectra at 86 K, as shown in Fig 13(B) illustrates the peak synthesis of the high intensity broad peak centered around 649.45 $cm^{-1}$.

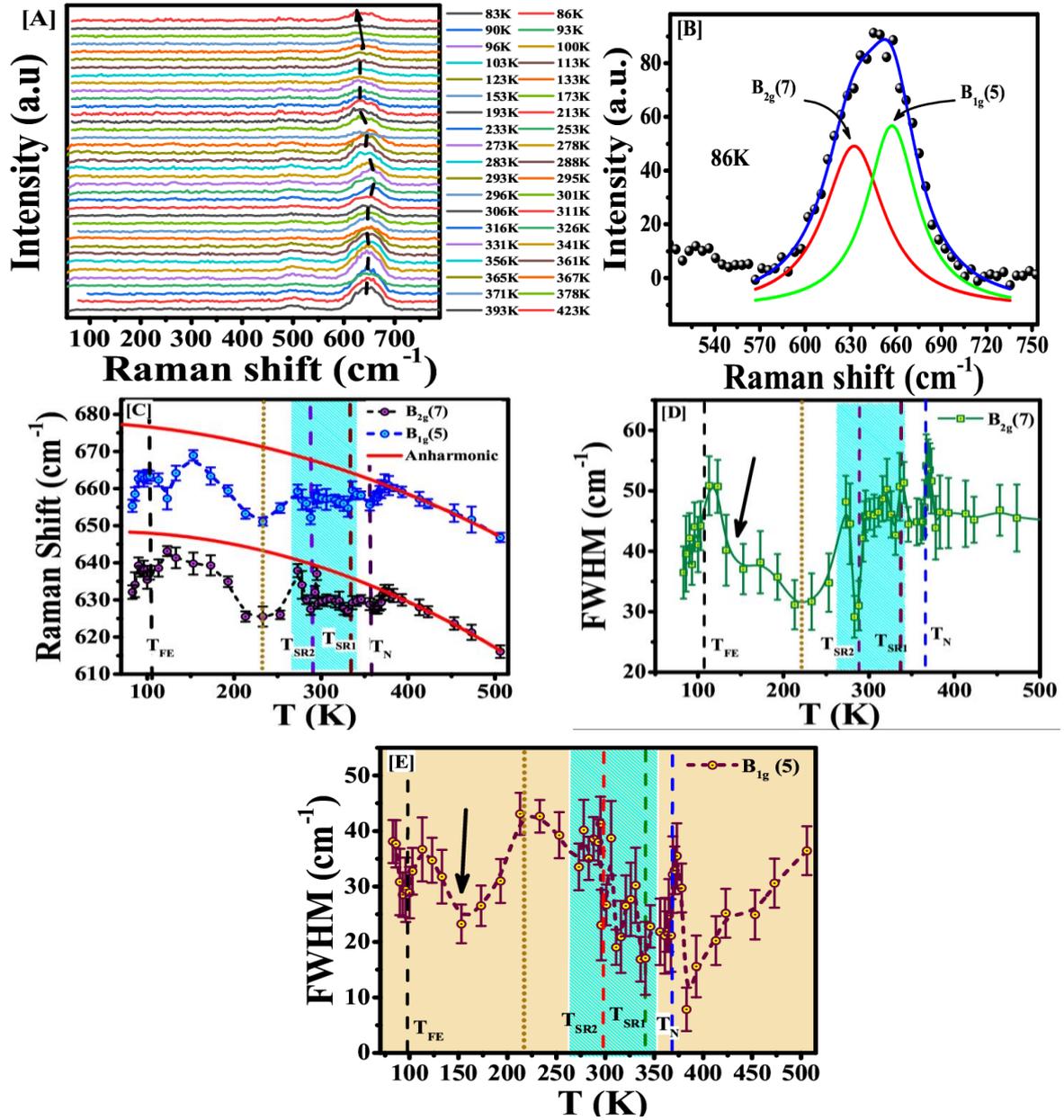

**Fig 13.** Panel (A) shows the unpolarised Raman spectra of SYFM (58-42) at several fixed temperatures. Panel (B) shows the $B_{2g}(7)$ and $B_{1g}(5)$ Raman modes at 86K. Panel (C) shows the T variation of the Raman shift of the $B_{2g}(7)$ and $B_{1g}(5)$ Raman modes between 83- 506K. The solid lines are phonon anharmonicity variation with T and broken lines are guide to the eye. Panel (D) and (E) shows the T variation of the FWHM of the $B_{2g}(7)$ and $B_{1g}(5)$ Raman modes between 83-506K. Panel (F) and (G) shows the enlarged portion of the plots of (D) and (E) respectively around the magnetic transitions. Broken lines with symbols are guide to the eye.

Its Lorentzian peak deconvolution yielded two synthetic peaks corresponding to M-O (M=Fe/Mn) octahedral stretching modes of $B_{2g}(7)$ and $B_{1g}(5)$ symmetries centered around 632.34±3.35 cm$^{-1}$ and 657.37± 2.13 cm$^{-1}$ respectively at 86 K. Most importantly, these modes couple with the atomic spins in several Mn doped systems and hence plays the key role in displaying spin-phonon coupling in these systems [36, 37]. As evidenced from Fig 13(C), the T variation of Raman shift (RS) of both $B_{2g}$ (7) and $B_{1g}$ (5) modes suffer significant softening at $T_N$ (blue dashed line in Fig 13(C)). The cubic anharmonic variation of the RS and linewidths (LW) with temperature are described as [70]:

$$\omega(T) = \omega(0) - A\left(1 + \frac{2}{e^{\hbar\omega_0/2K_BT} - 1}\right) \quad \ldots\ldots\ldots(12a)$$

$$\Gamma(T) = \Gamma(0) + B\left(1 + \frac{2}{e^{\hbar\omega_0/2K_BT} - 1}\right) \quad \ldots\ldots\ldots(12b)$$

where $\omega(0)$ and $\Gamma(0)$ are the intrinsic frequency and linewidth of the optical mode due to the defect respectively; A and B are coefficients for cubic anharmonic processes. It is clear from Fig 13(C) that Eq. 12(a) can satisfactorily describe T variation of the RS of the $B_{2g}$ (7) and $B_{1g}$ (5) modes [solid (red) lines] for T$\geq$ 373 K only. For T < 373 K, significant phonon softening occurs in both the Raman modes. With decreasing T, a dip appeared at $T_N$ causing the change in slope in T variation of the phonon frequency from usual anharmonicity. With further decreasing T similar features appeared at $T_{SR1}$ and $T_{SR2}$ as well.

This implies that **q** dependent spin-spin correlation arising from the phase coexistence anomalously affects the optical phonon modes. The competing SPC strengths from AFM/WFM correlations within the material at different temperatures can give rise to anomalous changes in Phonon mode frequencies line widths which is suggested in recent literatures also [73]. As the hence reveal the presence of significant spin-phonon coupling i.e phonon modulation of the spin-exchange integral in the system [71-73]. Similar convincing anomalies in the phonon mode frequencies are also evident around $T_{FE}$/ $T_{comp}$, indicates that lattice modulation at the *compensation temperature* is stabilizing the *ferroelectric* ground state in SYFM (58-42). This appearance of electric polarization in conjunction with the magnetization reversal convincingly suggest that SYFM (58-42) to be a Type II multiferroic [13, 26].

Fig 13(D) and 13(E) displays the T variation of the linewidths (FWHM) of the $B_{2g}$ (7) and $B_{1g}$ (5) modes respectively in the measured T range. Same convincing anomalous T variation of the linewidths of Raman modes are also evident at the magnetic and ferroelectric transitions. As shown in Fig 13(D) there is an apparent temperature independency of the linewidth of $B_{2g}$(7) mode far above $T_N$, that can be attributed to the competition between a decrease in the linewidths due to the absence of magnon–phonon interaction above $T_N$ and an increase in the line widths with increased site disordering with increasing temperature [72]. Unlike the $B_{2g}$ (7) mode the FWHM of the $B_{1g}$ (5) mode exhibited a monotonous increase above 373K due to anharmonicity as dictated by Eq. (12b). These anomalous features in RS and FWHM occurring at the magnetic and ferroelectric transitions confirms that SPC gives rise to the observed MD and ferroelectricity in the SYFM (58-42) [3]. Previous reports in cho et.al.[74] showed thermally stimulated depolarization currents (TSDC) is the actual origin of the pyroelectric current for T ~ 110 K in $YFe_{0.8}Mn_{0.2}O_3$ single crystals, thereby generalizing that no long range polar order of electric dipoles occur in $YFe_{0.6}Mn_{0.4}O_3$ [3]. However, in our present study, the substitution of Sm in high percentage at Y site induces a *true* ferroelectric state which is also in the vicinity of the *compensation point* through SPC. Hence we suggest the necessity to re-investigation several doped and undoped $RFeO_3$ systems such as $SmFeO_3$ [33], regarding occurrence of intrinsic electric polarization in conjunction with the NM state.

It is worth notable that there exist certain critical anomalies in the T variation of linewidths and the phonon frequencies of both the modes at 220 K [indicated as dashed yellow line in Fig 13 (C)-(E)] and around 153 K [indicated by arrow in Fig 13(D) and 13(E)]. Since the temperatures of these anomalies coincides with the slope changes in thermal variation of the overall resistivity and grain resistance as shown in Fig 11 (B) and 10 (G) respectively, it confirms the earlier suggestion of electron localization caused by electron-phonon interactions at 220 K.

## IV. Conclusion.

In conclusion, the polycrystalline samples of $Y_{0.5}Sm_{0.5}Fe_{0.58}Mn_{0.42}O_3$ below $T_N$ exhibits an incomplete second order spin reorientation transition at $T_{SR1}$ that is immediately followed by a first order spin reorientation at $T_{SR2}$, leading to completion of the spin reorientation in to a nearly collinear antiferromagnetic state. The delicate interplay between the $Sm^{3+}$- $Fe^{3+}/Mn^{3+}$ anisotropic exchange interaction and anisotropy *nature* of the $Mn^{3+}$ ions causes the two consecutive spin reorientation transitions below $T_N$ in the present system. Astonishingly a re-entrant spinglass like state have been observed for T below 70K and existing with the long range ordered magnetic phase in the material. Robust magnetodielectric effects can be observed at RT as illustrated from the magnetic field dependent dielectric constant that scales linearly to the squared magnetization in the high field regime for H > 1.5 T (at 300 K) as described by the Ginzburg-Landau theory. Significant spin phonon coupling is observed at $T_N$, $T_{SR1}$ and across FOMT, involving magnetoelectric coupling and ferroelectricity above liquid $N_2$ temperatures. All-over the present study reveals the intercorrelation of the physical properties of the material $Y_{0.5}Sm_{0.5}Fe_{0.58}Mn_{0.42}O_3$ reveals a delicate interplay of the spin, charge and lattice degrees of freedom that suggest the material to be a potential candidate for multifunctional applications.

## Supplementary Material

Supplementary material for combined frequency explicit plots of imaginary electric modulus and impedance plots modelling with equivalent circuits.

## Acknowledgement

We acknowledges UGC-DAE CSR Indore for measurement support. S.R. and S.P. also acknowledges Mr. Balram Thakur for helping in SQUID-VSM measurements in UGC-DAE CSR Kalapakkam Node. The whole work was supported by the Ministry of Human Resource and Development (MHRD), India, under the indigenous fellowship scheme. This work is also partially funded by project approved under collaborative research scheme (UGC DAE- CRS) vide project no.: CSR-IC-255/2017-18/1336 dated 31-03- 2018.

## Data Availability Statement

The data generated and/or analyzed during the current study are not publicly available for legal/ethical reasons but are available from the corresponding author on reasonable request.